\pdfoutput=1

\documentclass[11pt,twoside,a4paper,cmspaper,final,collab]{cms-tdr}
\def\svnVersion{481779P}\def\cmsCernNoTag{CERN-EP-2018-203}\def\cmsCernDate{\today}\def\cmsMessage{Published in Physical Review D as \href{http://dx.doi.org/10.1103/PhysRevD.98.092011}{\doi{10.1103/PhysRevD.98.092011.}}}

\begin{document}\cmsNoteHeader{EXO-17-018}

\hyphenation{had-ron-i-za-tion}
\hyphenation{cal-or-i-me-ter}
\hyphenation{de-vices}
\RCS$Revision: 478456 $
\RCS$HeadURL: svn+ssh://svn.cern.ch/reps/tdr2/papers/EXO-17-018/trunk/EXO-17-018.tex $
\RCS$Id: EXO-17-018.tex 478456 2018-10-17 22:16:45Z jchu $

\newlength\cmsFigWidth
\ifthenelse{\boolean{cms@external}}{\setlength\cmsFigWidth{0.85\columnwidth}}{\setlength\cmsFigWidth{0.4\textwidth}}
\ifthenelse{\boolean{cms@external}}{\providecommand{\cmsLeft}{upper\xspace}}{\providecommand{\cmsLeft}{left\xspace}}
\ifthenelse{\boolean{cms@external}}{\providecommand{\cmsRight}{lower\xspace}}{\providecommand{\cmsRight}{right\xspace}}
\ifthenelse{\boolean{cms@external}}{\providecommand{\NA}{\ensuremath{\cdots}\xspace}}{\providecommand{\NA}{\ensuremath{\text{---}}\xspace}}
\ifthenelse{\boolean{cms@external}}{\providecommand{\CL}{C.L.\xspace}}{\providecommand{\CL}{CL\xspace}}

\newlength\cmsTabSkip\setlength{\cmsTabSkip}{1ex}
\newcommand{\intlumiTotal}{38.5\fbinv}
\newcommand{\nsigmadxy}{\ensuremath{\abs{d_{xy}}/\sigma_{d_{xy}}}\xspace}
\newcommand{\dbv}{\ensuremath{d_{\mathrm{BV}}}\xspace}
\newcommand{\dvv}{\ensuremath{d_{\mathrm{VV}}}\xspace}
\newcommand{\dvvc}{\ensuremath{d_{\mathrm{VV}}^{\kern 0.15em\mathrm{C}}}\xspace}
\newcommand{\dphivv}{\ensuremath{\Delta\phi_{\mathrm{VV}}}\xspace}
\newcommand{\dphijj}{\ensuremath{\Delta\phi_{\mathrm{JJ}}}\xspace}

\cmsNoteHeader{EXO-17-018}

\title{Search for long-lived particles with displaced vertices in multijet events in proton-proton collisions at \texorpdfstring{$\sqrt{s}=13$\TeV}{sqrt(s)=13 TeV}}

\date{\today}

\abstract{
Results are reported from a search for long-lived particles in
proton-proton collisions at $\sqrt{s}=13\TeV$ delivered by the CERN
LHC and collected by the CMS experiment.  The data sample, which was
recorded during 2015 and 2016, corresponds to an integrated
luminosity of 38.5\fbinv.  This search uses benchmark signal models
in which long-lived particles are pair-produced and each decays into
two or more quarks, leading to a signal with multiple jets and two
displaced vertices composed of many tracks.  No events with two
well-separated high-track-multiplicity vertices are observed.  Upper
limits are placed on models of $R$-parity violating supersymmetry in
which the long-lived particles are neutralinos or gluinos decaying
solely into multijet final states or top squarks decaying solely
into dijet final states.  For neutralino, gluino, or top squark
masses between 800 and 2600\GeV and mean proper decay lengths
between 1 and 40\mm, the analysis excludes cross sections above
0.3\unit{fb} at 95\% confidence level.  Gluino and top squark masses
are excluded below 2200 and 1400\GeV, respectively, for mean proper
decay lengths between 0.6 and 80\mm.  A method is provided for
extending the results to other models with pair-produced long-lived
particles.
}

\hypersetup{
pdfauthor={CMS Collaboration},
pdftitle={Search for long-lived particles with displaced vertices in multijet events in proton-proton collisions at sqrt(s)=13 TeV},
pdfsubject={CMS},
pdfkeywords={CMS, physics, displaced vertices}}

\maketitle

\section{Introduction}

Many theories for physics beyond the standard model (SM) predict the
pair production of long-lived particles decaying to final states
with two or more jets.  Some examples include $R$-parity violating
(RPV) supersymmetry (SUSY)~\cite{Barbier:2004ez}, split
SUSY~\cite{Hewett:2004nw}, hidden valley
models~\cite{Strassler:2006im}, and weakly interacting massive
particle baryogenesis~\cite{Cui:2014twa}.  Searches for long-lived
particles significantly expand the parameter space of physics beyond
the SM probed by the experiments at the CERN LHC.

\begin{figure}[hbtp]
\centering
\includegraphics[width=0.48\textwidth]{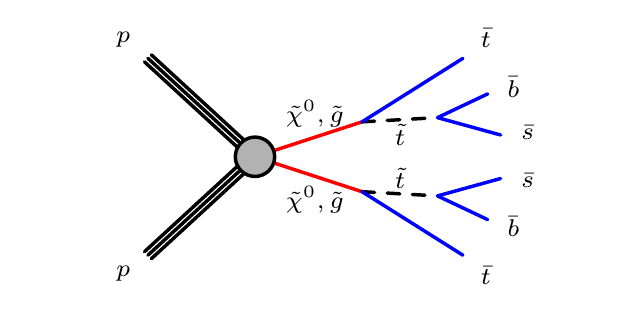}
\includegraphics[width=0.48\textwidth]{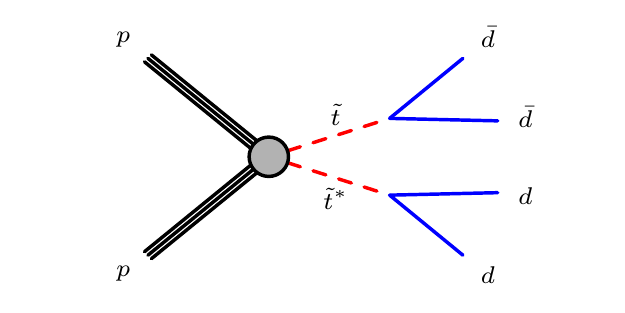}
\caption{Diagrams for the multijet (\cmsLeft) and dijet (\cmsRight)
benchmark signal models used in this analysis.  In the multijet
signal model, long-lived neutralinos (\PSGcz) or gluinos (\PSg)
decay into top, bottom, and strange antiquarks, via a virtual top
squark (\PSQt).  In the dijet signal model, long-lived top squarks
decay into two down antiquarks.  The charge conjugate processes are
also considered.}
\label{fig:diagrams}
\end{figure}

This analysis is sensitive to models of new physics in which pairs
of long-lived particles decay to final states with multiple charged
particles.  We present results for two benchmark signal models, as
well as a method for applying the results more generally.  The
``multijet'' benchmark signal is motivated by a minimal flavor
violating model of RPV SUSY~\cite{Yuval} in which the lightest SUSY
particle is a neutralino or gluino, either of which is produced in
pairs.  The neutralino or gluino is long-lived and decays into a top
antiquark and a virtual top squark, and the virtual top squark
decays into strange and bottom antiquarks, resulting in a final
state with many jets.  The ``dijet'' benchmark signal corresponds to
an RPV phenomenological model in which pair-produced long-lived top
squarks each decay into two down antiquarks~\cite{Csaba}.  The
diagrams for the multijet and dijet signal models are shown in
Fig.~\ref{fig:diagrams}.

The experimental signature of long-lived exotic particle pairs is
two displaced vertices, each consisting of multiple charged-particle
trajectories intersecting at a single point.  In this analysis, a
custom vertex reconstruction algorithm identifies displaced vertices
in the CMS detector.  We focus on signals with intermediate
lifetimes, corresponding to mean proper decay lengths $c\tau$ from
0.1 to 100\mm, by identifying vertices that are displaced from the
beam axis but within the radius of the beam pipe.  The signal is
distinguished from the SM background based on the separation between
the vertices: signal events have two well-separated vertices, while
background events are dominated by events with only one displaced
vertex, usually close to the beam axis.

The CMS Collaboration searched for displaced vertices in
proton-proton ($\Pp\Pp$) collisions at a center-of-mass energy of
$\sqrt{s}=8$\TeV in 2012~\cite{CMS-SUS-14-020}.  This analysis is an
updated version of the search, using $\Pp\Pp$ collisions collected
at $\sqrt{s}=13$\TeV.  It improves upon the previous analysis,
because of better background suppression along with a refined
procedure for estimating the background and the associated
systematic uncertainties.  A similar analysis was performed by the
ATLAS Collaboration~\cite{ATLAS-SUSY-2016-08}.  The ATLAS, CMS, and
LHCb Collaborations have also searched for displaced jets or
leptons~\cite{CMS-B2G-12-024,CMS-EXO-12-037,CMS-EXO-12-038,ATLAS-EXOT-2012-28,ATLAS-EXOT-2013-12,LHCb-PAPER-2016-014,LHCb-PAPER-2016-047,LHCb-PAPER-2016-065,CMS-EXO-16-003},
displaced photons~\cite{ATLAS-SUSY-2013-17}, and displaced lepton
jets~\cite{ATLAS-EXOT-2013-22}.  The analysis reported here is
sensitive to shorter lifetimes than those probed by previous
analyses.

\section{The CMS detector}

The central feature of the CMS detector is a superconducting
solenoid providing a magnetic field of 3.8\unit{T} aligned with the
proton beam direction.  Contained within the field volume of the
solenoid are a silicon pixel and strip tracker, a lead tungstate
electromagnetic calorimeter, and a brass and scintillator hadron
calorimeter.  Muon tracking chambers are embedded in the steel
flux-return yoke that surrounds the solenoid.  A more detailed
description of the CMS detector, together with a definition of the
coordinate system used and the relevant kinematic variables, can be
found in Ref.~\cite{Chatrchyan:2008zzk}.

The silicon tracker, which is particularly relevant to this
analysis, measures the trajectories of charged particles in the
range of pseudorapidity, $\eta$, up to $\abs{\eta} < 2.5$.  For
nonisolated particles with transverse momentum, \pt, of 1 to 10\GeV
and $\abs{\eta} < 1.4$, the track resolutions are typically 1.5\% in
\pt, 25--90\mum in the impact parameter in the transverse plane, and
45--150\mum in the impact parameter in the longitudinal
direction~\cite{TRK-11-001}.  Jets are reconstructed from
particle-flow~\cite{CMS-PRF-14-001} candidates using the anti-\kt
algorithm~\cite{Cacciari:2008gp, Cacciari:2011ma} with a distance
parameter of 0.4.

Events of interest are selected using a two-tiered trigger
system~\cite{Khachatryan:2016bia}.  The first level is composed of
custom hardware processors, and the second level consists of a farm
of processors running a version of the full event reconstruction
software optimized for fast processing.

\section{Event samples}

The data sample used in this analysis corresponds to a total
integrated luminosity of \intlumiTotal, collected in $\Pp\Pp$
collisions at $\sqrt{s}=13\TeV$ in 2015 and 2016.  Events are
selected using a trigger initially requiring $\HT > 800\GeV$, where
\HT is the scalar sum of the \pt of jets in the event with $\pt >
40\GeV$.  In the last data-taking period of 2016, corresponding to
22\% of the total integrated luminosity, the higher instantaneous
luminosity required the \HT threshold to be raised to 900\GeV.

Simulated events are used to model the signal processes.  In the
multijet and dijet signal models, long-lived particles are produced
in pairs; the ``multijet'' and ``dijet'' refer to the decay of each
long-lived particle.  For the multijet signals, the long-lived
particle is a neutralino that undergoes a three-body decay into top,
bottom, and strange quarks.  In this analysis, the final results are
the same if the neutralinos are replaced with gluinos.  For the
dijet signals, the long-lived particle is a top squark that decays
into two down antiquarks.  Signal samples with various neutralino or
top squark masses $m$ ($300 \leq m \leq 2600\GeV$) and lifetimes
$\tau$ ($0.1 \leq c\tau \leq 100\mm$) are produced using \PYTHIA
8.212~\cite{PYTHIA} with the NNPDF2.3QED parton distribution
functions~\cite{Ball:2013hta}.

Backgrounds arising from SM processes are dominated by multijet and
top quark pair production (\ttbar) events.  The multijet processes
include \PQb quark pair production events.  The multijet and \ttbar
events are simulated using \MGvATNLO 2.2.2~\cite{MADGRAPH} with the
NNPDF3.0 parton distribution functions~\cite{Ball:2014uwa}, at
leading order with MLM merging~\cite{Alwall:2007fs} for the multijet
events and at next-to-leading order with FxFx
merging~\cite{Frederix:2012ps} for the \ttbar events.

For all samples, hadronization, showering, and $R$-hadron physics
are simulated using \PYTHIA 8.212.  The underlying event tunes used
are CUETP8M1~\cite{CMS-GEN-14-001} for the signal samples and the
multijet background samples, and
CUETP8M2T4~\cite{CMS-PAS-TOP-16-021} for the \ttbar samples.  The
detector response for all simulated samples is modeled using a
\GEANTfour-based simulation~\cite{GEANT} of the CMS detector.  The
effects of additional $\Pp\Pp$ interactions within the same or
nearby bunch crossings (``pileup'') are included by overlaying
additional simulated minimum-bias events, such that the resulting
distribution of the number of interactions per bunch crossing
matches that observed in the experiment.

\section{Event preselection}
\label{sec:eventsel}

For an event to be selected for further analysis, it must have at
least four jets, each with $\pt > 20\GeV$ and $\abs{\eta} < 2.5$.
Since the final states for the signal models considered all have at
least four quarks, this requirement has little impact on signal
events but is beneficial in suppressing background.

To ensure that the efficiency of the \HT trigger is well understood,
a stricter requirement of $\HT > 1000\GeV$ is applied offline, where
\HT is the scalar sum of the \pt of jets with $\pt > 40\GeV$, to
match the trigger jet definition.  For events with at least four
jets and $\HT > 1000\GeV$, the trigger efficiency, determined using
events satisfying a trigger requiring the presence of at least one
muon, is $(99 \pm 1)\%$.

\section{Vertex reconstruction and selection}

Displaced vertices are reconstructed from tracks in the silicon
tracker.  These tracks are required to have $\pt > 1\GeV$;
measurements in at least two layers of the pixel detector, including
one in the innermost layer; measurements in at least six layers of
the strip detector if $\abs{\eta} < 2$, or in at least seven layers
if $\abs{\eta} \geq 2$; and significance of the impact parameter
with respect to the beam axis measured in the $x$-$y$ plane (the
magnitude of the impact parameter divided by its uncertainty,
referred to as \nsigmadxy) of at least 4.  The first three criteria
are track quality requirements, imposed in order to select tracks
with small impact parameter uncertainties.  The requirement on track
\nsigmadxy favors vertices that are displaced from the beam axis.

The vertex reconstruction algorithm forms seed vertices from all
pairs of tracks satisfying the track selection criteria, and then
merges them iteratively until no track is used more than once.  A
set of tracks is considered to be a vertex if a fit with the Kalman
filter approach~\cite{Kalman} has a $\chi^2$ per degree of freedom
($\chi^2/\mathrm{dof}$) that is less than 5.  Subsequently, for each
pair of vertices that shares a track, the vertices are merged if the
three-dimensional distance between the vertices is less than 4 times
the uncertainty in that distance and the fit has
$\chi^2/\mathrm{dof} < 5$.  Otherwise, the shared track is assigned
to one of the vertices depending on the value of its
three-dimensional impact parameter significance with respect to each
of the vertices: if both values are less than 1.5, the shared track
is assigned to the vertex that has more tracks already; if either
value is greater than 5, the shared track is dropped from that
vertex; otherwise, the shared track is assigned to the vertex with
respect to which it has a smaller impact parameter significance.  If
a track is removed from a vertex, that vertex is refit, and if the
fit satisfies the requirement of $\chi^2/\mathrm{dof} < 5$, the old
vertex is replaced with the new one; otherwise it is dropped
entirely.

This procedure produces multiple vertices per event, only some of
which are signal-like.  In order to select vertices with high
quality, we impose additional requirements: each vertex is required
to have at least five tracks; a distance from the detector origin
measured in the $x$-$y$ plane of less than 20\mm, to avoid vertices
from interactions in the beam pipe or detector material; a distance
from the beam axis measured in the $x$-$y$ plane, defined as \dbv,
of at least 0.1\mm, to suppress displaced primary vertices; and an
uncertainty in \dbv of less than 25\mum, to select only
well-reconstructed vertices.  The requirement on the uncertainty in
\dbv also suppresses displaced vertices from single \PQb jets, which
are composed of tracks that are mostly aligned with the vertex
displacement from the beam axis and have small opening angles
between the tracks.

Since signal events contain a pair of long-lived particles, we
require events to have two or more vertices satisfying the above
requirements.  The signal region is composed of these two-vertex
events.  Simulation predicts there is on the order of 1 background
event in the signal region for \intlumiTotal of data.  However,
establishing the possible presence of a signal relies on an accurate
determination of the background, and for this we rely on data.

The vertex selection requires each vertex to have five or more
tracks, but events with vertices with three or four tracks provide
valuable control samples.  These control samples, which are used to
test the background prediction, have a factor of 10--100 more
background events than in the signal region and negligible potential
signal contamination.  Simulation studies show that events
containing 3-track, 4-track, and $\geq$5-track vertices have similar
distributions of event variables, such as \HT, number of jets, and
quark flavor composition, as well as vertex variables, such as \dbv,
uncertainty in \dbv, and angular separation between tracks.

\section{Search strategy}
\label{sec:searchstrategy}

The signal is discriminated from the SM background using the
distance between the two vertices measured in the $x$-$y$ plane,
which is defined as \dvv.  In signal events, the two long-lived
particles are emitted approximately back-to-back, leading to large
separations.  If an event has more than two vertices, the two
vertices with the highest number of tracks are selected for the \dvv
calculation.  In the case in which two vertices have the same number
of tracks, the vertex with the higher mass is chosen, where the mass
is reconstructed using the momenta of the tracks associated with the
vertex, assuming that the particles associated with the tracks have
the mass of a charged pion.

\begin{figure}[hbtp]
\centering
\includegraphics[width=0.48\textwidth]{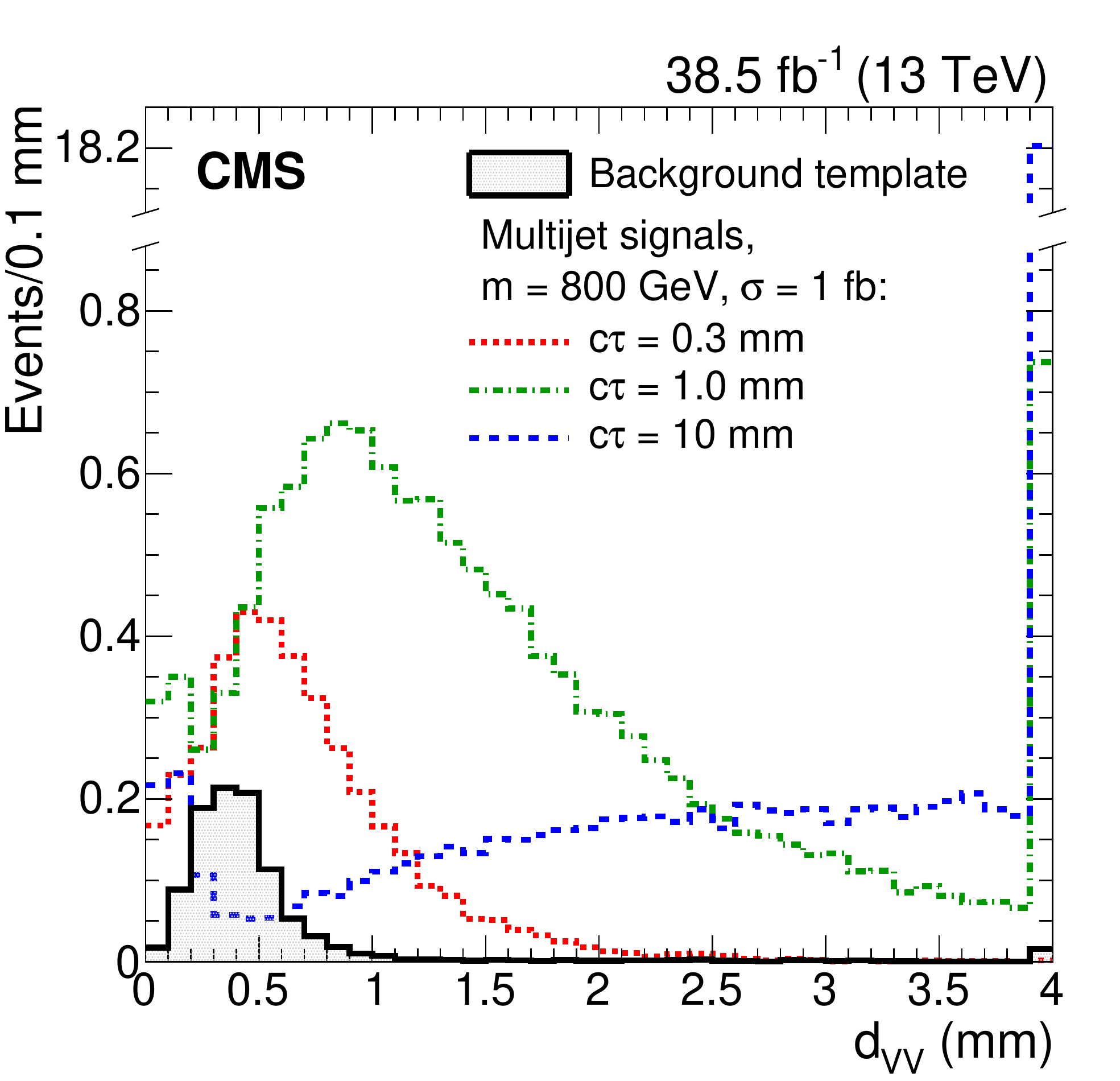}
\caption{Distribution of the distance between vertices in the
$x$-$y$ plane, \dvv, for simulated multijet signals with $m =
800\GeV$, production cross section 1\unit{fb}, and $c\tau = 0.3$,
1.0, and 10\mm, with the background template overlaid.  All vertex
and event selection criteria have been applied.  The last bin
includes the overflow events.}
\label{fig:templates}
\end{figure}

We fit the distribution of \dvv to extract the signal, using
templates to represent the \dvv distributions for signal and
background.  The signal \dvv templates are taken directly from
simulation, with a distinct template for each signal mass and
lifetime.  The background template is constructed from events in
data that have exactly one vertex, as described in
Section~\ref{sec:bkgest}.  Figure~\ref{fig:templates} shows examples
of the \dvv distribution for simulated multijet signals with $m =
800\GeV$ and production cross section 1\unit{fb}, with the
background template overlaid.  The distributions depend primarily on
the signal lifetime; those for other signal masses and for the dijet
signals are similar.  The small peaks at low values of \dvv are
associated with events for which the two vertices are reconstructed
from the same long-lived particle, with the effect being larger for
the multijet signals.

In the signal extraction procedure, the \dvv distribution is broken
into three bins: 0--0.4\mm, 0.4--0.7\mm, and 0.7--40\mm.  The two
bins with $\dvv > 0.4\mm$ have low background.  This division
maximizes the signal significance for scenarios with intermediate
and long lifetimes.

\begin{figure}[hbtp]
\centering
\includegraphics[width=0.48\textwidth]{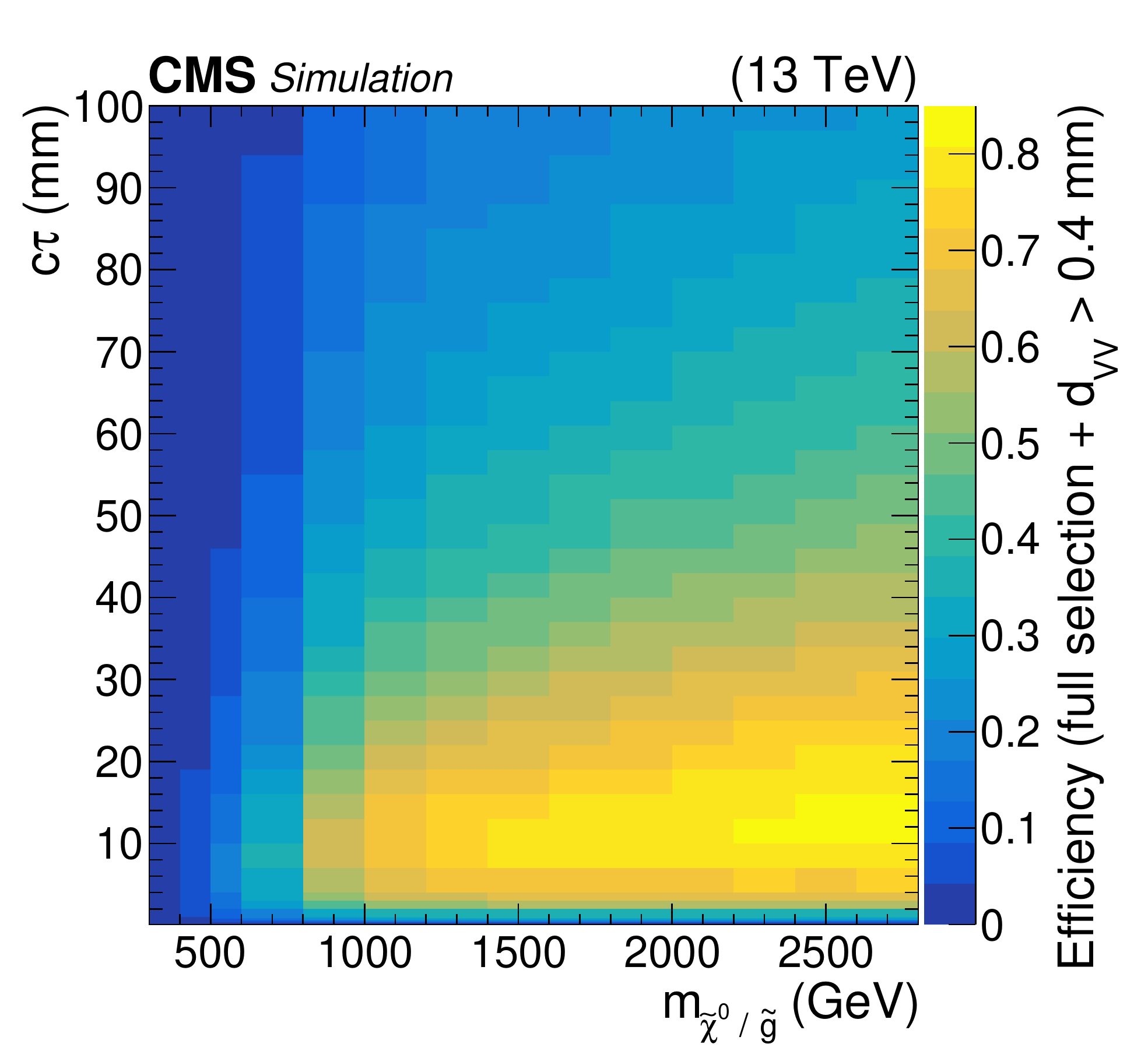}
\includegraphics[width=0.48\textwidth]{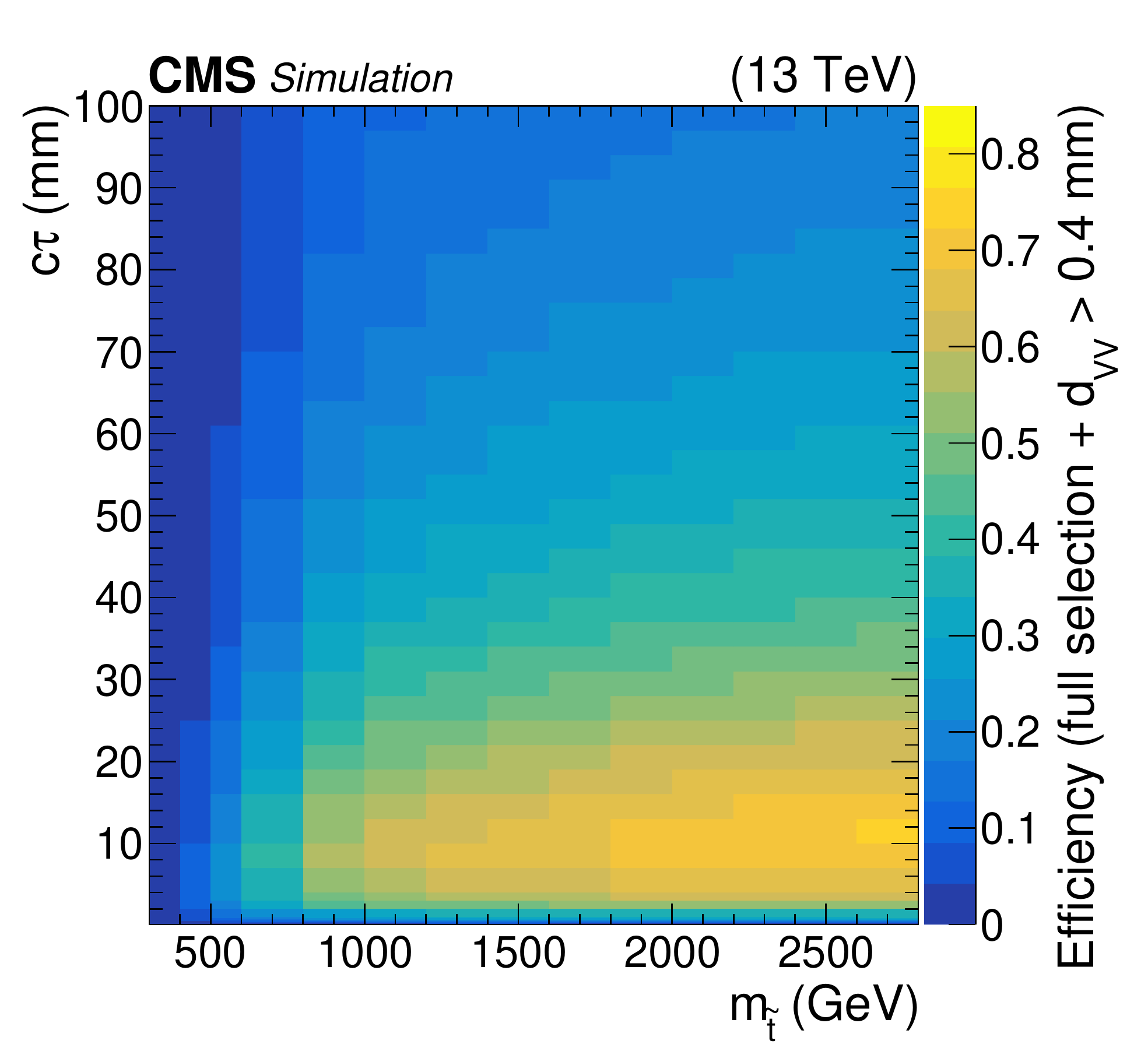}
\caption{Signal efficiency as a function of signal mass and
lifetime, for the multijet (\cmsLeft) and dijet (\cmsRight) signal
samples.  All vertex and event selection criteria have been applied,
as well as the requirement $\dvv > 0.4\mm$.}
\label{fig:sigeff}
\end{figure}

Figure~\ref{fig:sigeff} shows the signal efficiency as a function of
signal mass and lifetime in the region $\dvv > 0.4\mm$.  The signal
efficiency increases with increasing mass because the events are
more likely to satisfy the \HT trigger requirement.  As lifetime
increases, the signal efficiency initially increases because of
better separation from the beam axis, but then starts to decrease
when the lifetime is so long that decays occur more often beyond the
fiducial limit at the beam pipe.  The efficiency is above 10\% for
$c\tau > 0.4\mm$ and $m > 800\GeV$.

\section{Background template}
\label{sec:bkgest}

Displaced vertices in background events arise from one or more
misreconstructed tracks overlapping with other tracks.  These events
are dominated by multijet and \ttbar processes.  The tracks can
arise from light parton or \PQb quark jets, with those from \PQb
quark decays typically producing slightly larger vertex
displacements.  Displaced vertices composed of tracks from a single
\PQb quark jet are rejected because of the vertex requirement on the
uncertainty in \dbv.  Background events with two vertices arise from
coincidences of misreconstructed vertices, whose displacements are
independent apart from small correlations due to events with \PQb
quark pairs.  Accordingly, we construct the two-vertex background
template, denoted by \dvvc, by combining information from events in
data that have exactly one vertex, and then correcting for possible
correlations between vertices.  There are approximately 1000 times
more events with only one vertex than there are with two or more
vertices, consistently for 3-track, 4-track, and $\geq$5-track
vertices.  Table~\ref{tab:yields} lists the number of events in each
of the event categories.

\begin{table}[htbp!]
\centering
\topcaption{Event yields in data.  The ``one-vertex'' events have
exactly one vertex with the specified number of tracks, and the
``two-vertex'' events have two or more vertices each with the
specified number of tracks.  The control samples are composed of the
events with 3-track and 4-track vertices, the background template is
constructed using the $\geq$5-track one-vertex events, and the
signal region consists of the $\geq$5-track two-vertex events.}
\begin{scotch}{lcccc}
Event category & 3-track & 4-track $\times$ 3-track & 4-track & $\geq$5-track \\
\hline
one-vertex     &  109090 &                      \NA &   11923 &          1183 \\
two-vertex     &     478 &                       99 &       7 &             1 \\
\end{scotch}
\label{tab:yields}
\end{table}

Each entry in the \dvvc template is calculated from two values of
\dbv and a value of \dphivv, where \dbv is the distance measured in
the $x$-$y$ plane from the beam axis to one vertex, and \dphivv is
the azimuthal angle between the two vertices.  The template also
includes corrections for the merging of nearby vertices in the
vertex reconstruction algorithm and for possible correlations
between individual vertices in background events with pairs of \PQb
quarks.  The following paragraphs describe each of the inputs to the
\dvvc template construction method.

\begin{figure}[hbtp]
\centering
\includegraphics[width=0.48\textwidth]{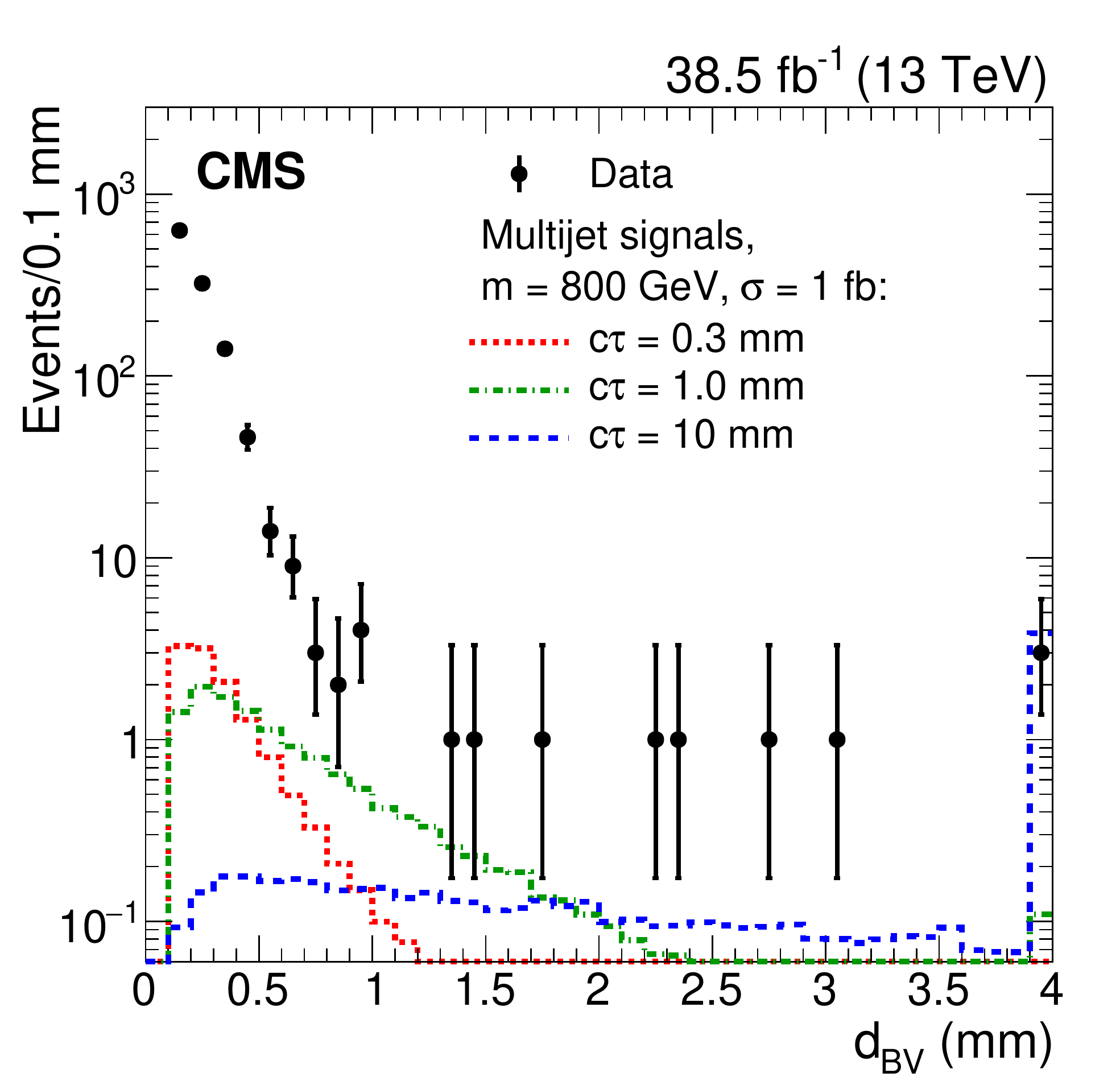}
\caption{Distribution of \dbv in $\geq$5-track one-vertex events for
data and simulated multijet signals with $m = 800\GeV$, production
cross section 1\unit{fb}, and $c\tau = 0.3$, 1.0, and 10\mm.  Event
preselection and vertex selection criteria have been applied.  The
last bin includes the overflow events.}
\label{fig:dbv}
\end{figure}

The \dbv values are sampled from the distribution shown in
Fig.~\ref{fig:dbv} for the $\geq$5-track one-vertex events in data.
The distribution starts at 0.1\mm because of the fiducial
requirement imposed to avoid primary vertices, and falls off
exponentially.  Signal contamination in the one-vertex sample is
negligible for values of the signal cross section that have not been
excluded by the previous similar analysis~\cite{CMS-SUS-14-020}.

The statistical uncertainty in the \dvvc template, taken as the
root-mean-square of yields in an ensemble of simulated pseudodata
sets, depends on the number of entries in the parent \dbv
distribution.  To ensure sufficient sampling of the tail of this
distribution, the number of entries in the \dvvc template is 20
times the number of one-vertex events.

Values of \dphivv are approximated by sampling the distribution of
jets in data.  Since background vertices arise from misreconstructed
tracks in jets, their position vectors tend to be correlated with
jet momentum vectors.  The angle between vertex positions can
therefore be modeled using the observed distribution of azimuthal
angles between pairs of jets, denoted as \dphijj.  The \dphijj
distribution used for the \dvvc construction is taken from the
3-track one-vertex sample, which has a greater number of events than
the 4-track and $\geq$5-track one-vertex samples.  There are no
significant differences in the \dphijj distribution among these
three samples.

To emulate the behavior of the vertex reconstruction algorithm in
merging overlapping vertices, the \dvvc template is corrected by the
survival probability of pairs of vertices as a function of \dvv.
This efficiency is estimated by counting the number of remaining
vertex pairs at each iteration of the vertex reconstruction
algorithm.  The efficiency correction suppresses small \dvvc values,
resulting in a yield in the first \dvvc bin that is lower by a factor
of approximately 2.

Pair production of \PQb quarks introduces \dbv correlations in
two-vertex events that are not accounted for when pairing single
vertices at random.  This is because the tracks from \PQb quark
decays are more likely to satisfy the track \nsigmadxy requirement
and therefore produce vertices.  In simulation, the mean \dbv in
events with \PQb quarks is higher than in events without \PQb quarks
by $47 \pm 1\mum$ for 3-track vertices, by $52 \pm 3\mum$ for
4-track vertices, and by $50 \pm 6\mum$ for $\geq$5-track vertices.
The fractions of events with \PQb quarks are consistent across the
3-track, 4-track, and $\geq$5-track vertex samples: approximately
50\% in one-vertex events and approximately 78\% in two-vertex
events.  We determine corrections to the \dvvc template for these
\dbv correlations by constructing \dvvc separately for simulated
background events with and without generated \PQb quarks, combining
the distributions in the ratio of two-vertex events with and without
\PQb quarks, and then dividing the resulting distribution by the
nominal \dvvc template.  The \PQb quark correction enhances larger
\dvvc values, resulting in a yield in the last \dvvc bin that is
higher by a factor of $1.6 \pm 0.4$.

\begin{figure*}[hbtp]
\centering
\includegraphics[width=0.48\textwidth]{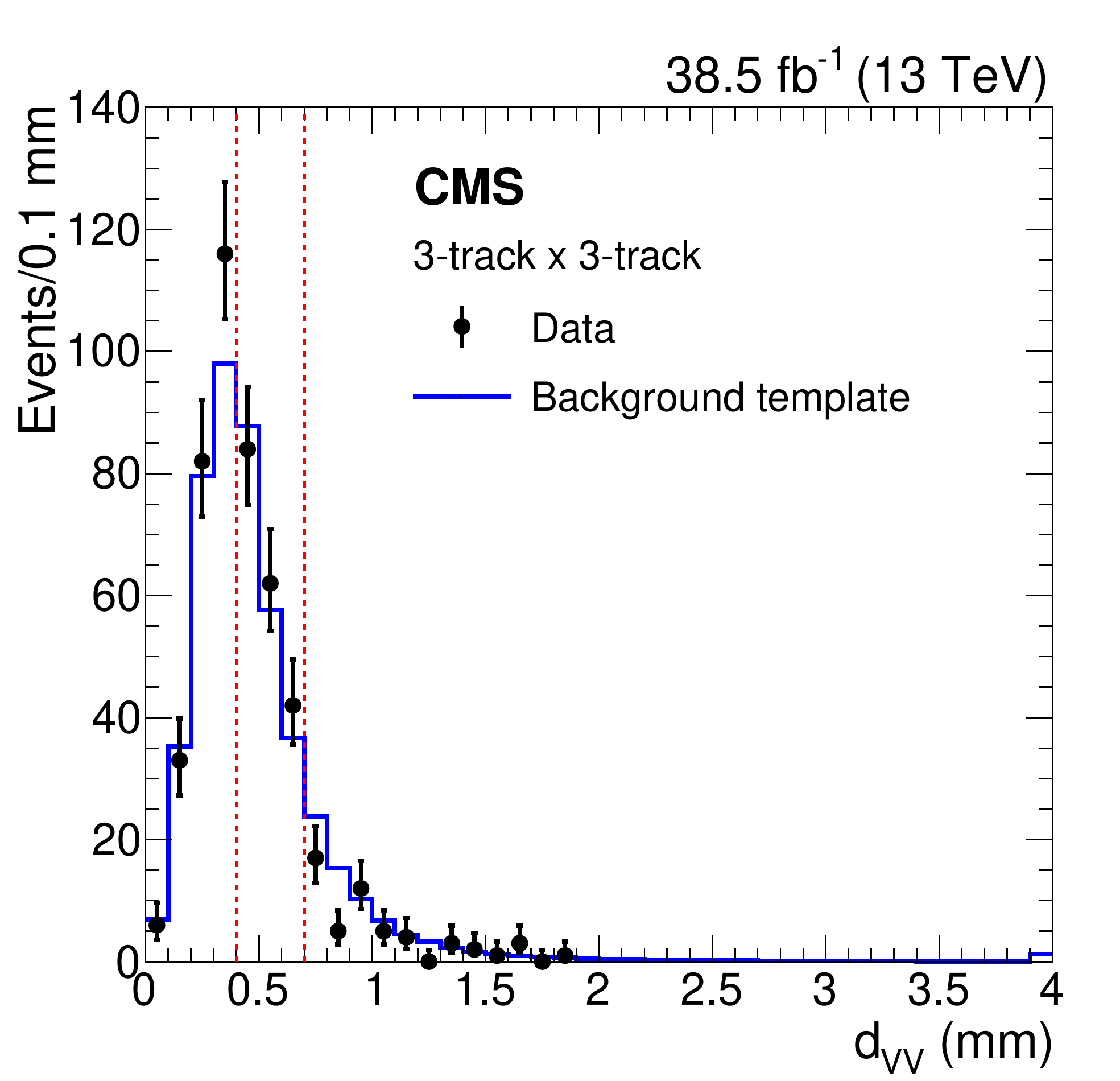}
\includegraphics[width=0.48\textwidth]{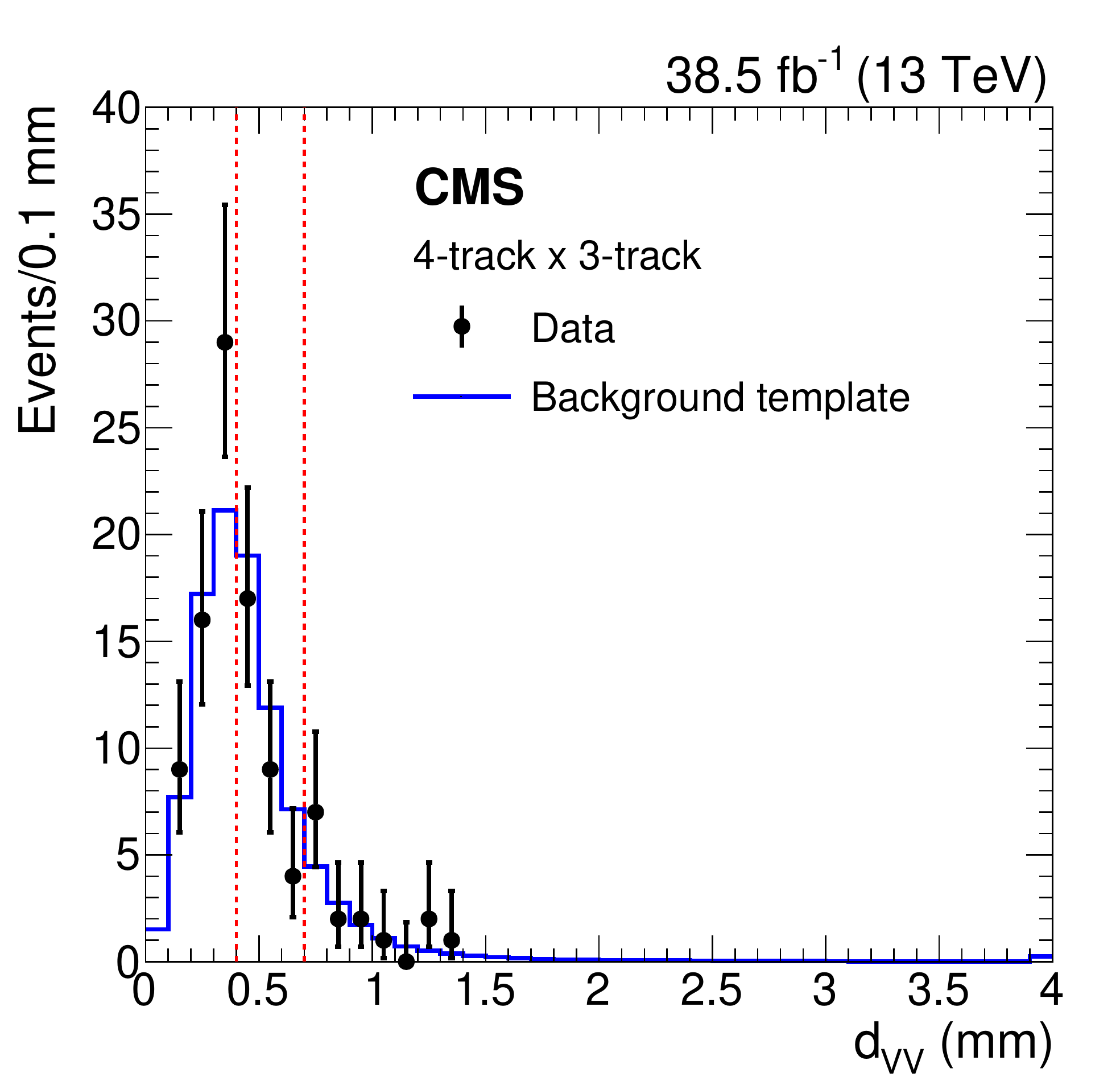}
\includegraphics[width=0.48\textwidth]{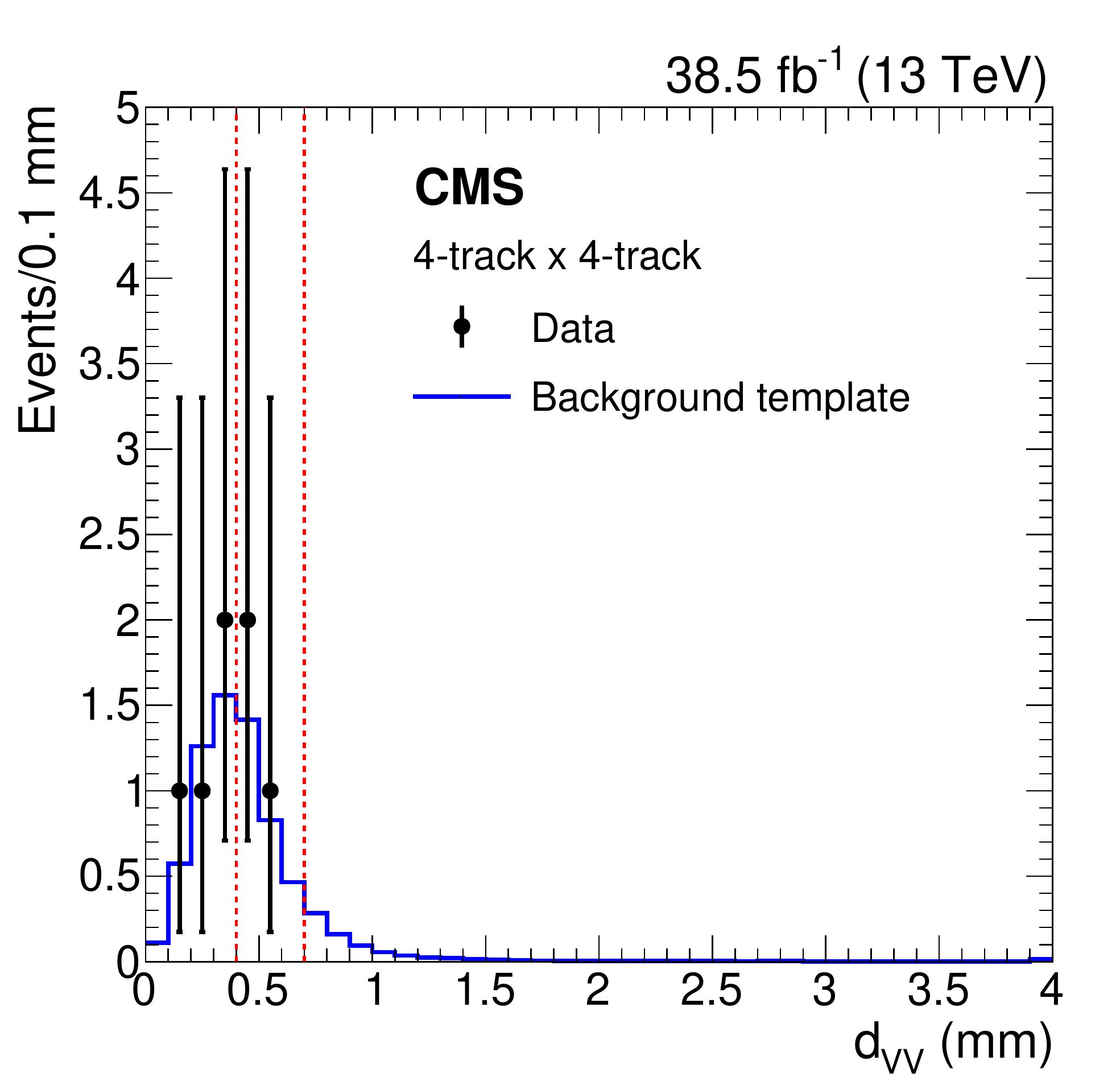}
\includegraphics[width=0.48\textwidth]{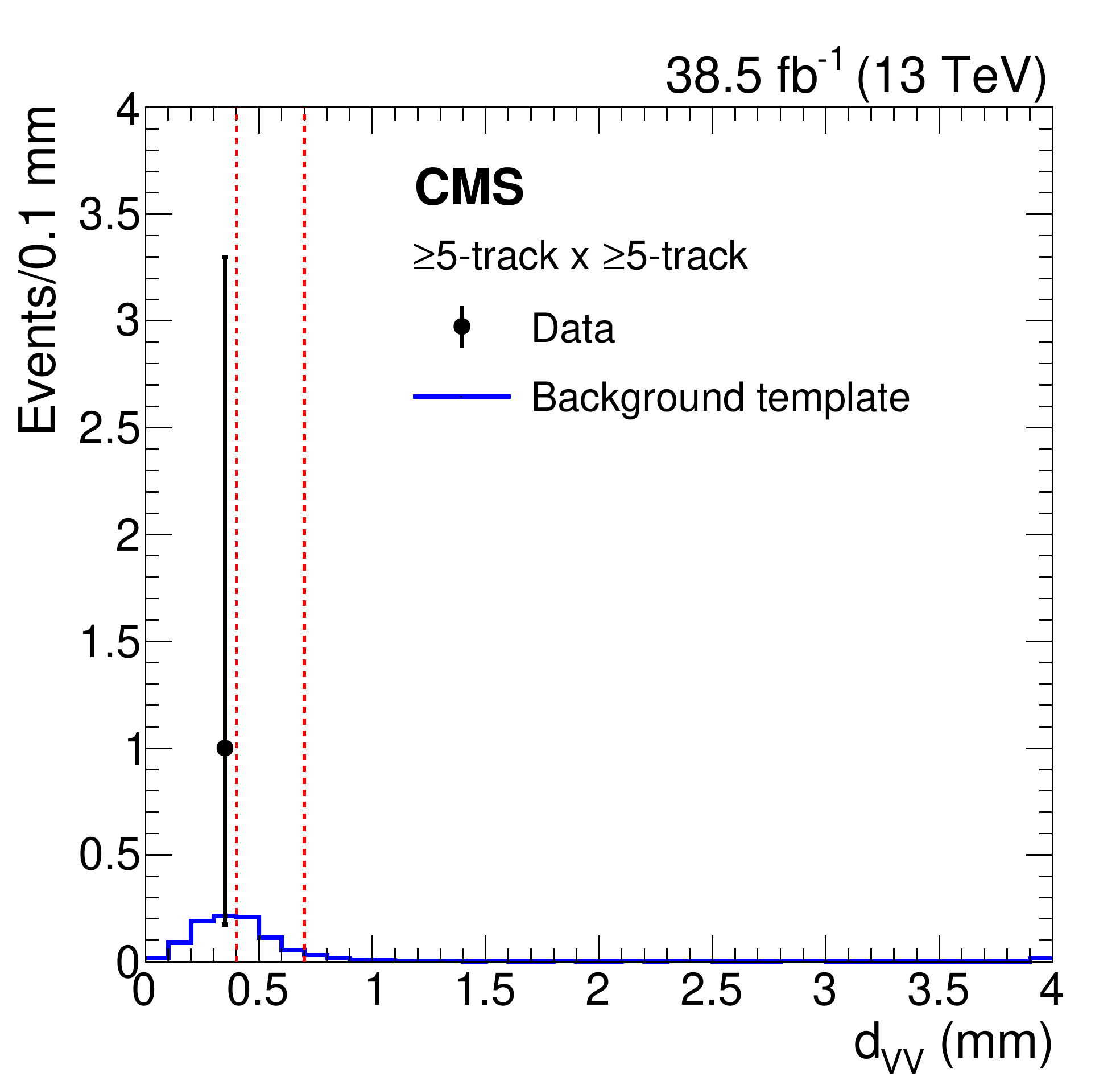}
\caption{Distribution of the distance between vertices in the
$x$-$y$ plane in two-vertex events.  The points show the data
(\dvv), and the solid lines show the background template (\dvvc)
normalized to the data, for events with two 3-track vertices (upper
left), one 4-track vertex and one 3-track vertex (upper right), two
4-track vertices (lower left), and two $\geq$5-track vertices (lower
right).  In each plot, the last bin includes the overflow events.
The dotted lines indicate the boundaries between the three bins used
in the fit.}
\label{fig:closure}
\end{figure*}

Evidence that the background template construction method is valid
is presented in the upper left, upper right, and lower left plots in
Fig.~\ref{fig:closure}, where \dvvc is compared to the observed
two-vertex \dvv distributions in the low-track-multiplicity control
samples in data.  There is good agreement between the relative \dvvc
and \dvv populations in each of the three bins of the final fit.
For example, in the 3-track control sample, where this agreement is
most stringently tested, the ratios \dvvc/\dvv are $0.93 \pm 0.06$
in the 0--0.4\mm bin, $0.97 \pm 0.07$ in the 0.4--0.7\mm bin, and
$1.44 \pm 0.20$ in the 0.7--40\mm bin.

The background template for the signal region is shown in the lower
right plot in Fig.~\ref{fig:closure}.

\section{Systematic uncertainties}

The signal yield is extracted from a fit of the signal and
background templates to the observed \dvv distribution.  The free
parameters are the normalizations of the signal and background
templates, subject to the constraint that their combined yield
matches the data.  The result of the fit relies on the relative
yields in the three bins of the templates, but is insensitive to the
fine details of the template distributions.  This section describes
the systematic uncertainties in the background template.  It also
addresses the systematic uncertainties in the signal efficiencies
and templates.

\subsection{Systematic uncertainties in signal efficiencies and templates}
The signal \dvv templates are taken directly from simulation of
benchmark models with clearly specified parameters, so the
systematic uncertainties arise from biases in the detector
simulation and reconstruction.  The dominant source of uncertainty
is due to the vertex reconstruction efficiency.  Smaller effects
arise from track resolution, pileup, jet energy scale and
resolution, integrated luminosity, and trigger efficiency.

The effect due to the vertex reconstruction efficiency is evaluated
by comparing the efficiency in data and simulation to reconstruct
signal-like vertices created by displacing tracks artificially.  In
events passing the preselection requirements
(Section~\ref{sec:eventsel}), we choose some number of light parton
and \PQb quark jets that have $\pt > 50\GeV$, $\abs{\eta} < 2.5$,
and at least four particle-flow candidates.  We then artificially
displace the tracks associated with those jets as described below.

The magnitude of the displacement vector is sampled from an
exponential distribution with scale parameter $c\tau = 10\mm$,
restricted to values between 0.3 and 20\mm.  The direction of the
displacement vector is calculated from the vector sum of the
momentum of the jets.  This direction is smeared to emulate the
difference between the vertex displacement direction and jet
momentum direction in signal events due to mismeasurements from
tracking inefficiency and missing neutral particles.

The track selection requirements and vertex reconstruction algorithm
are applied to the resulting set of tracks.  We then evaluate the
fraction of events in which a vertex satisfying all vertex selection
requirements is reconstructed near the artificial displacement
position.  This one-vertex reconstruction efficiency is evaluated for different numbers of
displaced light parton or \PQb quark jets.  The largest disagreement
between data and simulation gives an 11.5\% relative uncertainty in the one-vertex efficiency,
implying a 23\% relative uncertainty in the two-vertex efficiency.  Varying the scale
parameter of the exponential distribution or the amount that the
direction is smeared within reasonable values has negligible effect
on the difference between data and simulation.

The difference in vertex reconstruction efficiency between data and
simulation could also depend on the magnitude of the artificial
displacement.  This dependence is found to be small, and the
resulting difference in the signal \dvv templates has a negligible
effect on the signal yield extracted from the fit.

The selection of the tracks used in the vertex reconstruction
requires that each track has a value of \nsigmadxy of at least 4.
The efficiency of this requirement is sensitive to the impact
parameter resolution of the tracks.  The mean impact parameter
uncertainty is 2\% larger in data than in simulation.  The magnitude
of this effect is quantified by tightening the requirement on the
transverse impact parameter significance by 2\% and evaluating the
change in the signal efficiency.  The maximum effect on the various
signal masses and lifetimes, 5\%, is taken to be the systematic
uncertainty in the signal efficiency.  This effect is corrected for
in the vertex resolution study discussed earlier.

The uncertainties in the jet energy scale and
resolution~\cite{Khachatryan:2016kdb} could affect the total jet
energy and change the probability that events satisfy the \HT
selection.  Varying the jet energy scale by one standard deviation
results in a change in the signal efficiency of 5\% or less for all
signal samples, and varying the jet energy resolution by one
standard deviation changes the efficiency by 2\% or less.  We
therefore assign these as the corresponding systematic uncertainties
in the signal efficiency.

The uncertainty in the integrated luminosity is 2.3\% for
2015~\cite{CMS-PAS-LUM-15-001} and 2.5\% for
2016~\cite{CMS-PAS-LUM-17-001}.  The uncertainty in the signal
efficiency due to pileup is 2\%.  The uncertainty in the trigger
efficiency is 1\%.

Table~\ref{tab:sigeffsyst} summarizes the systematic uncertainties
in the signal efficiency.  We assume there are no correlations among
them, and add them in quadrature to obtain the overall uncertainty.

\begin{table}[htbp!]
\centering
\topcaption{Relative systematic uncertainties in the signal efficiency.  The
overall uncertainty is the sum in quadrature of the individual
uncertainties, assuming no correlations.}
\begin{scotch}{lc}
Systematic effect           & Uncertainty (\%) \\
\hline
Vertex reconstruction       & 23 \\
Track resolution            & 5 \\
Jet energy scale/resolution & 5 \\
Integrated luminosity       & 3 \\
Pileup                      & 2 \\
Trigger efficiency          & 1 \\
[\cmsTabSkip]
Overall                     & 24 \\
\end{scotch}
\label{tab:sigeffsyst}
\end{table}

\subsection{Systematic uncertainties in the background template}
The \dvvc background template is constructed from the large sample
of events in data with exactly one vertex.  Systematic uncertainties
in the background template arise from effects that could cause
differences between the constructed \dvvc distribution and the true
\dvv distribution of two-vertex background events.  The 3-track
control sample is used to evaluate the scale of these differences.
The deviation from unity of the ratio of the predicted yield in each
bin of the \dvvc template to the observed yield in the same bin,
which is referred to as the closure, is a measure of the systematic
uncertainty.  Additional uncertainties arise from effects that could
compromise the validity of applying the 3-track control sample to
the $\geq$5-track sample.

We check the assumption that closure of the \dvvc construction
method in 3-track vertices implies closure in $\geq$5-track vertices
by varying the inputs to the template construction procedure and
evaluating the resulting shifts in the \dvvc template.  Constructing
\dvvc involves sampling two values of \dbv and an angle between
vertices \dphivv, the efficiency to keep pairs of vertices as a
function of \dvv, and the \PQb quark correction factors.  Therefore,
the main effects are related to these distributions.  We include
additional systematic uncertainties to account for possible
differences in \dvvc predictions due to variations in these
distributions from 3-track vertices to $\geq$5-track vertices.

In background template construction, the \dphivv distribution is
modeled using the \dphijj distribution in 3-track one-vertex
events.  The \dphijj distribution in $\geq$5-track one-vertex events
is indistinguishable from that in 3-track one-vertex events.
Potential bias could arise if the distribution of angles between
jets and vertices differ for 3-track and $\geq$5-track vertices.
Indeed, the correlation between vertex displacement directions and
jet directions is smaller for $\geq$5-track vertices than for
3-track vertices.  To probe the impact, we construct \dvvc using a
variation of the \dphivv input in which we assume that the
displacement directions are uncorrelated with the jet momentum
directions and draw \dphivv from a uniform distribution.  We then
assign the fractional change in the \dvvc prediction in each bin as
the systematic uncertainty.

The template also depends on the probability that pairs of nearby
vertices will both survive the vertex reconstruction algorithm as a
function of their separation \dvv.  The efficiency to merge pairs of
vertices is determined from the vertex reconstruction algorithm.  To
assess the uncertainty due to variations in this efficiency, we use
a variation of the algorithm in which the seed vertices are composed
of five tracks, rather than the usual two.  We then construct a
variation of \dvvc using the resulting efficiency curve and take the
fractional change in the \dvvc prediction in each bin as the
systematic uncertainty.

The corrections to the \dvvc template that account for \dbv
correlations due to the pair production of \PQb quarks are derived
using the fraction of simulated 3-track two-vertex events with \PQb
quarks.  This fraction could differ for $\geq$5-track two-vertex
events.  To assess the related systematic uncertainty, we recompute
the \PQb quark corrections using the extreme case in which all
two-vertex events contain \PQb quarks, and determine the fractional
shifts in the \dvvc yields in each bin.

The statistical uncertainties in the \PQb quark corrections are also
taken as systematic uncertainties in the template.

The systematic uncertainty in the background template, \dvvc, is
estimated using a combination of the closure of the construction
method in the control sample of 3-track vertices and the difference
in effects from 3-track vertices to $\geq$5-track vertices.
Table~\ref{tab:bkgestsyst} lists the shifts arising from these
components for each of the three \dvv bins, along with their
statistical uncertainties.  The statistical uncertainties in the
shifts take into account the correlation between the default
template and the variation.  In assessing the overall systematic
uncertainty in the background template, we add in quadrature the
shifts and their uncertainties, assuming no correlations.

\begin{table*}[htbp!]
\centering
\topcaption{Systematic shifts in the background prediction in each
\dvvc bin arising from varying the construction of the \dvvc
template.  The overall systematic uncertainty is the sum in
quadrature of the shifts and their statistical uncertainties,
assuming no correlations among the sources.}
\begin{scotch}{lr@{ $\pm$ }lr@{ $\pm$ }lr@{ $\pm$ }l}
                                                      & \multicolumn{6}{c}{Shift (\%)}             \\
Systematic effect                                     & \multicolumn{2}{c}{0--0.4\mm} & \multicolumn{2}{c}{0.4--0.7\mm} & \multicolumn{2}{c}{0.7--40\mm} \\
\hline
Closure in 3-track control sample                     & $- 7$ & $6$ & $- 3$ & $7$ & $+44$ & $20$   \\
Difference from 3-track to $\geq$5-track vertices:    & \multicolumn{6}{c}{}                       \\
\hspace{0.1in} Modeling of \dphivv                    & $+ 4$ & $0$ & $- 5$ & $1$ & $- 2$ & $ 3$   \\
\hspace{0.1in} Modeling of vertex survival efficiency & $+20$ & $1$ & $-19$ & $2$ & $-26$ & $ 7$   \\
\hspace{0.1in} Modeling of \PQb quark correction      & $-11$ & $1$ & $+ 9$ & $2$ & $+18$ & $ 9$   \\
\PQb quark correction statistical uncertainty         & \multicolumn{2}{c}{$\pm3$} & \multicolumn{2}{c}{$\pm9$} & \multicolumn{2}{c}{$\pm36$} \\
[\cmsTabSkip]
Overall systematic uncertainty                        & \multicolumn{2}{c}{25} & \multicolumn{2}{c}{25} & \multicolumn{2}{c}{69} \\
\end{scotch}
\label{tab:bkgestsyst}
\end{table*}

\section{Signal extraction and statistical interpretation}

To determine the signal yield, we perform binned shape fits of the
signal and background templates to the \dvv distribution using an
extended likelihood method~\cite{Barlow}.

The background template is constructed from the one-vertex events in
data, while the signal templates are produced directly using the
\dvv distributions from simulation. There is one signal template for
each signal model, mass, and lifetime.

The lower right plot in Fig.~\ref{fig:closure} compares the \dvvc
and \dvv distributions in the signal region.  The observed number of
events in each bin, along with the predictions from the
background-only fit and from example signal models, are listed in
Table~\ref{tab:bincounts}.  The background-only fit normalizes the
prediction from the \dvvc background template to the observed number
of two-vertex events.  For the signal-plus-background fits, the
signal yield is constrained to be nonnegative.  Since there is only
one two-vertex event in the data, falling in the 0--0.4\mm \dvv bin,
the fits to the observed distribution prefer zero signal yield.

\begin{table*}[hbtp!]
\centering
\topcaption{For each \dvv bin in $\geq$5-track two-vertex events:
the predicted background yield from the background-only fit, the
observed yield, and the predicted signal yields for simulated
multijet signals with $m = 2000\GeV$, production cross section
1\unit{fb}, and $c\tau = 0.3$, 1.0, and 10\mm.  The systematic
uncertainties in the predicted background yields reflect the
fractional systematic uncertainties given in
Table~\ref{tab:bkgestsyst}, and the uncertainties in the predicted
signal yields reflect the fractional systematic uncertainty given in
Table~\ref{tab:sigeffsyst}.}
\begin{scotch}{rccr@{ $\pm$ }lr@{ $\pm$ }lr@{ $\pm$ }l}
             &                                    &          & \multicolumn{6}{c}{Predicted multijet signal yields} \\
$\dvv$ range & Fitted background yield            & Observed & \multicolumn{2}{c}{0.3\mm} & \multicolumn{2}{c}{1.0\mm} & \multicolumn{2}{c}{10\mm} \\
\hline
0--0.4\mm    & $0.51 \pm 0.01\stat \pm 0.13\syst$ & 1        & $2.8$ & $0.7$ & $ 3.5$ & $0.8$ & $ 1.0$ & $0.2$ \\
0.4--0.7\mm  & $0.37 \pm 0.02\stat \pm 0.09\syst$ & 0        & $2.0$ & $0.5$ & $ 3.7$ & $0.9$ & $ 0.5$ & $0.1$ \\
0.7--40\mm   & $0.12 \pm 0.02\stat \pm 0.08\syst$ & 0        & $1.1$ & $0.3$ & $11  $ & $3  $ & $31  $ & $7  $ \\
\end{scotch}
\label{tab:bincounts}
\end{table*}

Upper limits on the signal cross section are set using a Bayesian
technique~\cite{CowanPDGStat}.  A uniform prior is taken for
positive values of the signal cross section.  The signal efficiency
is constrained by a log-normal prior with a width of 24\%,
reflecting the overall uncertainty in the signal efficiency
(Table~\ref{tab:sigeffsyst}).  The only assumed uncertainty in the
shape of the signal templates is that due to the finite number of
events in the simulation; this uncertainty can be as large as 20\%
for the lower lifetime and mass samples that have small
efficiencies.  For the uncertainty in the background, log-normal
priors are taken for the yield in each bin, with widths given by the
fractional uncertainties listed in Table~\ref{tab:bkgestsyst}.

\begin{figure*}[hbtp]
\centering
\includegraphics[width=0.48\textwidth]{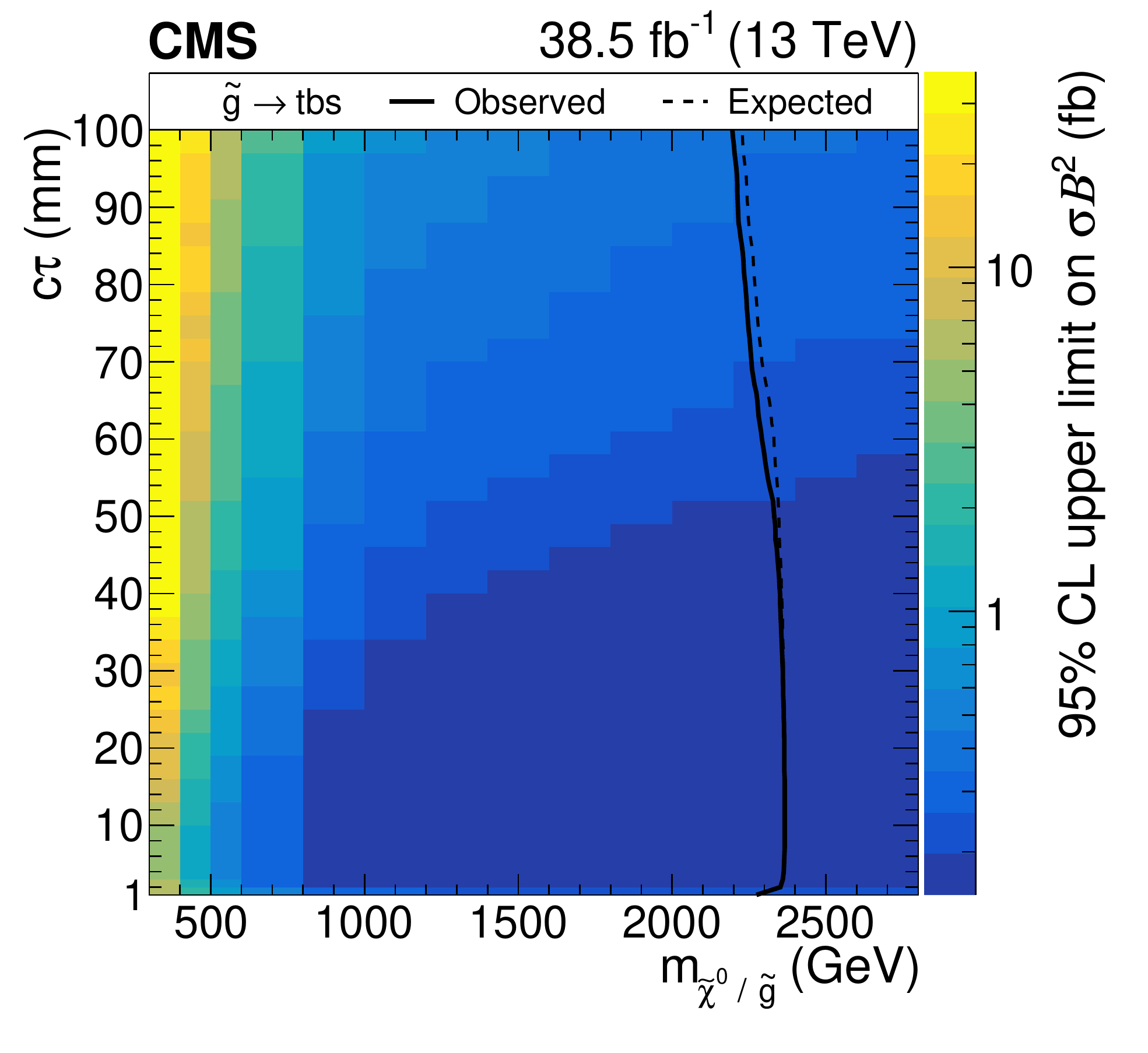}
\includegraphics[width=0.48\textwidth]{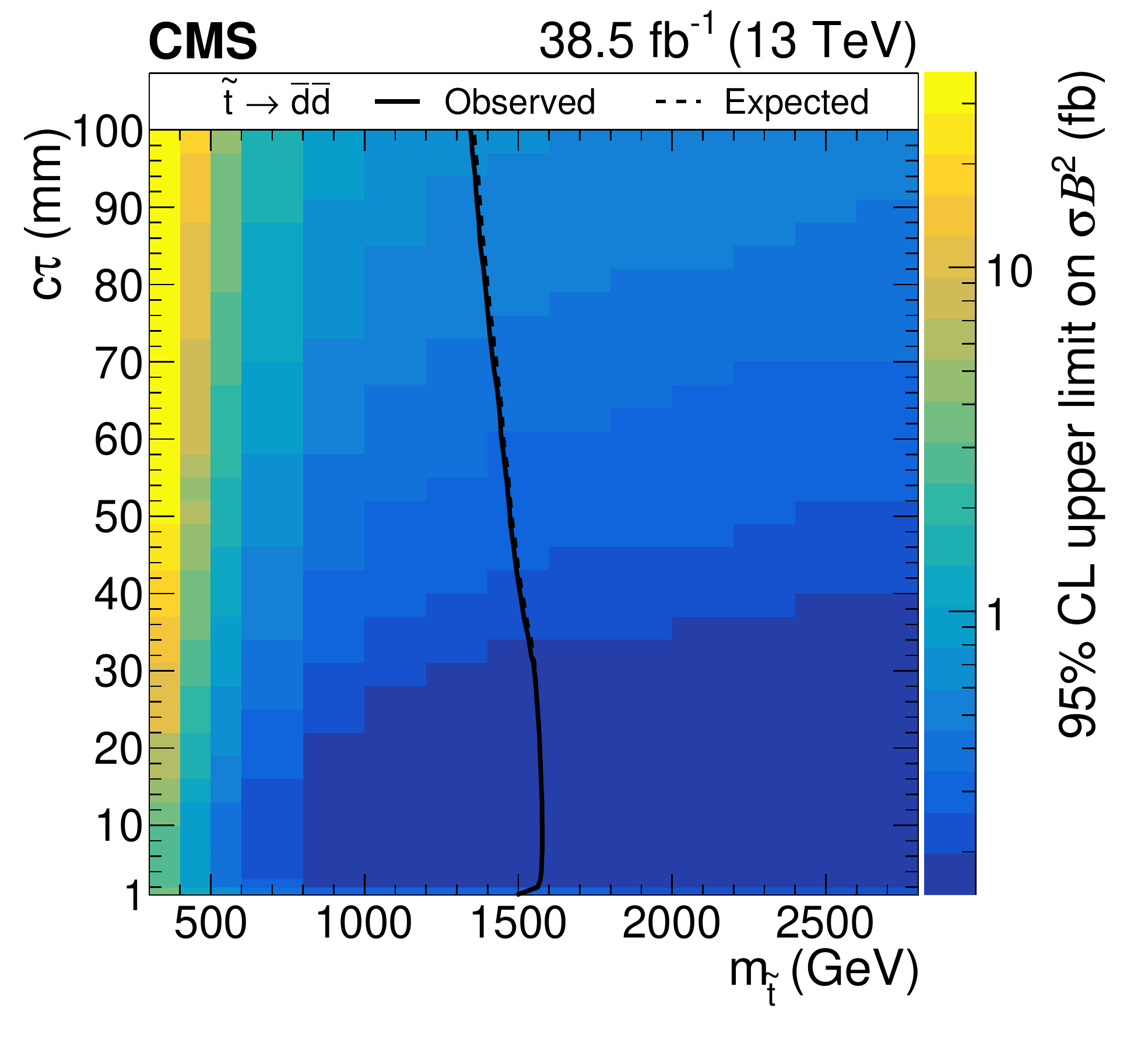}
\includegraphics[width=0.48\textwidth]{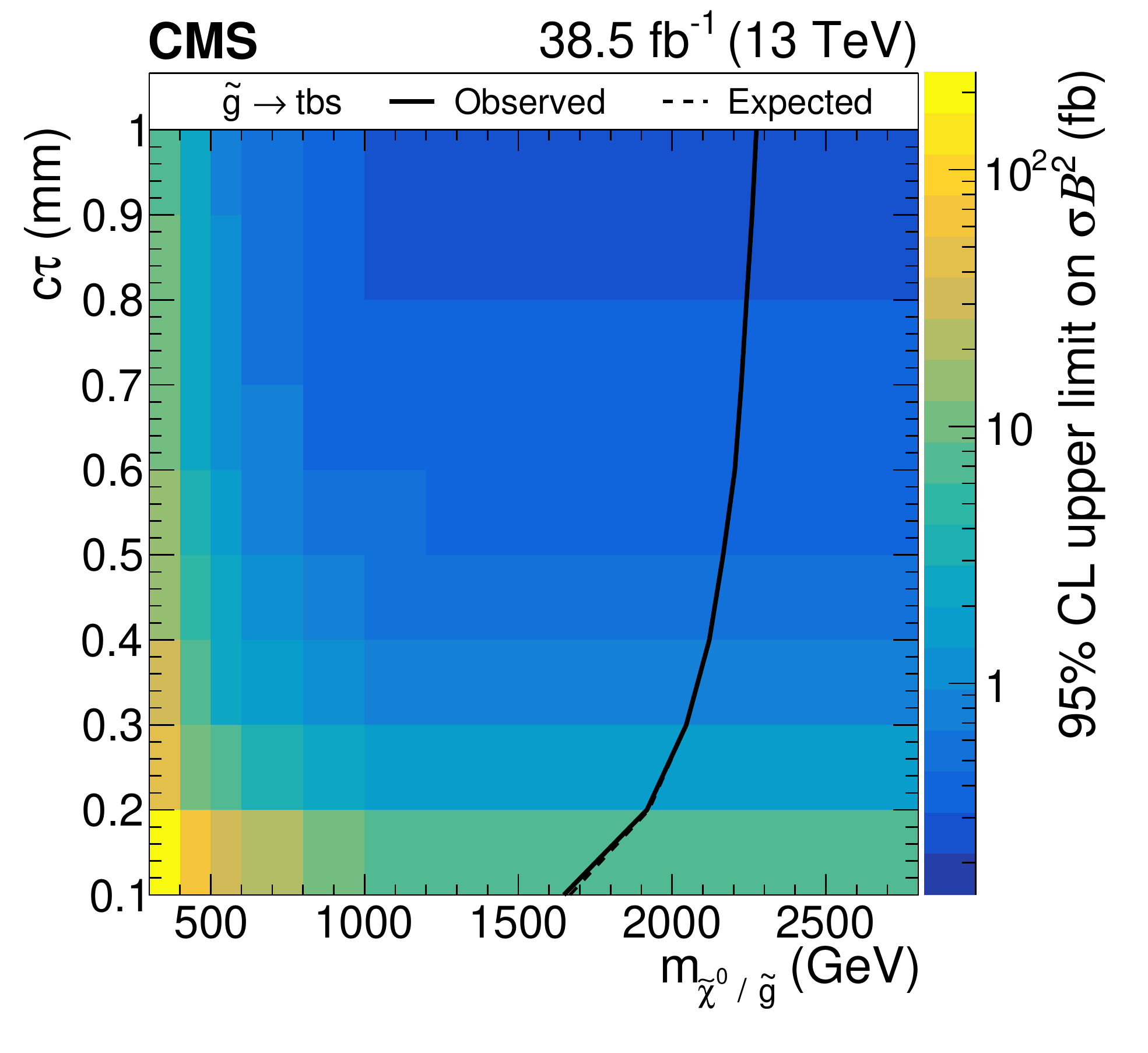}
\includegraphics[width=0.48\textwidth]{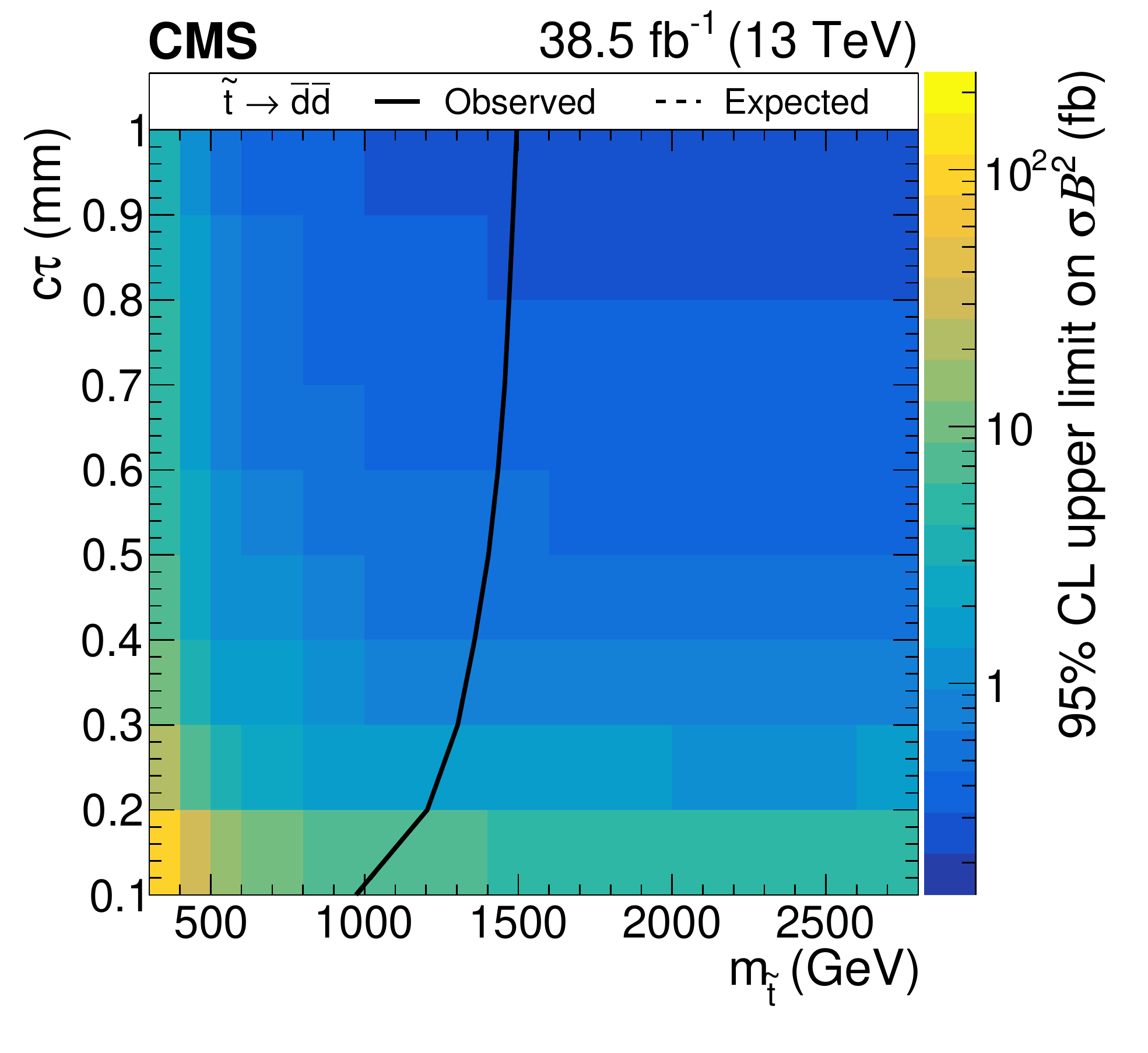}
\caption{Observed 95\% \CL upper limits on $\sigma\mathcal{B}^2$ for
the multijet (left) and dijet (right) signals as a function of mass
and mean proper decay length.  The upper plots span $c\tau$ from 1
to 100\mm, and the lower plots span $c\tau$ from 0.1 to 1\mm.  The
overlaid mass exclusion curves assume gluino pair production cross
sections for the multijet signals and top squark pair production
cross sections for the dijet signals, and 100\% branching fraction.}
\label{fig:limits}
\end{figure*}

\begin{figure*}[hbtp]
\centering
\includegraphics[width=0.45\textwidth]{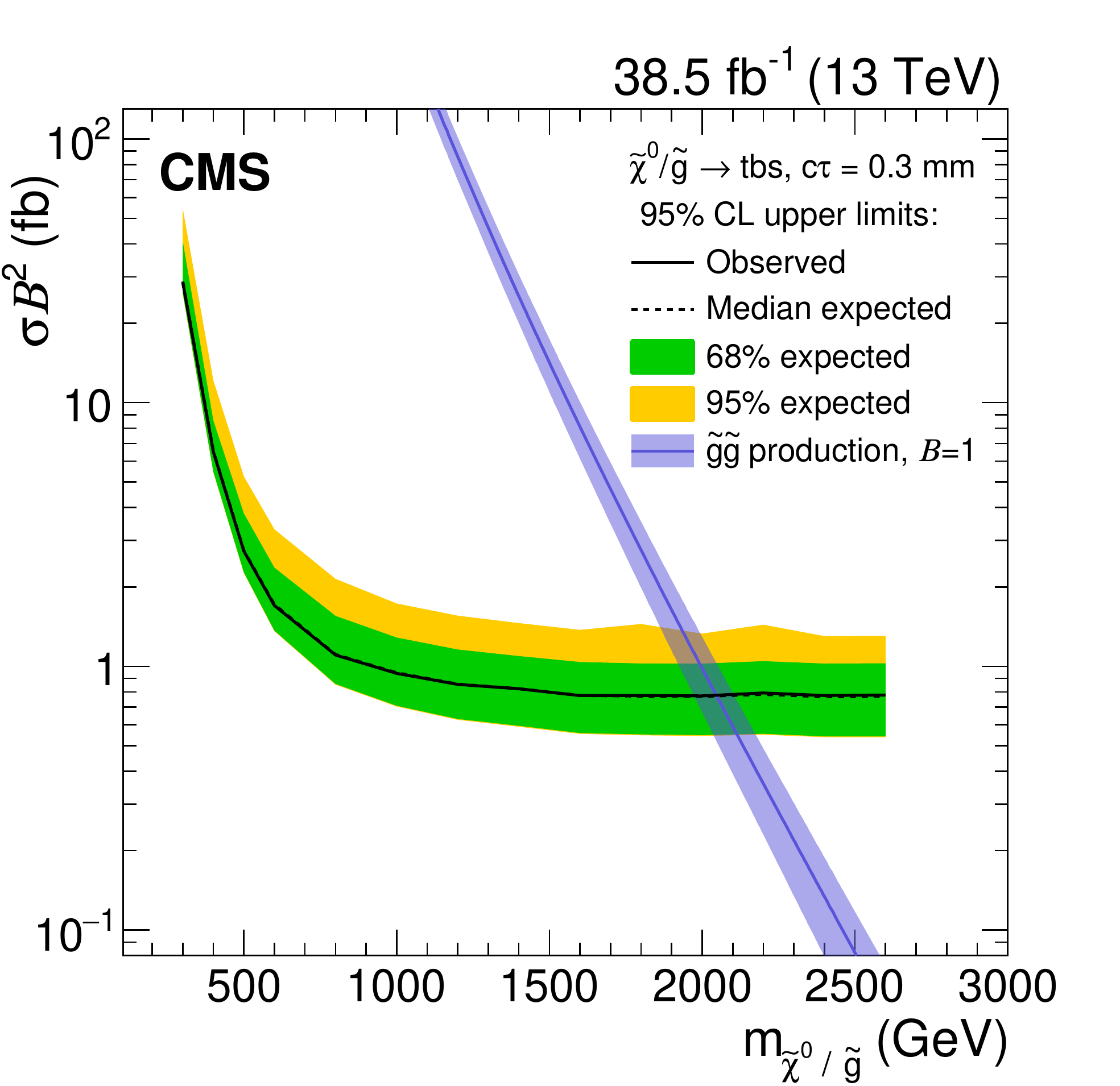}
\includegraphics[width=0.45\textwidth]{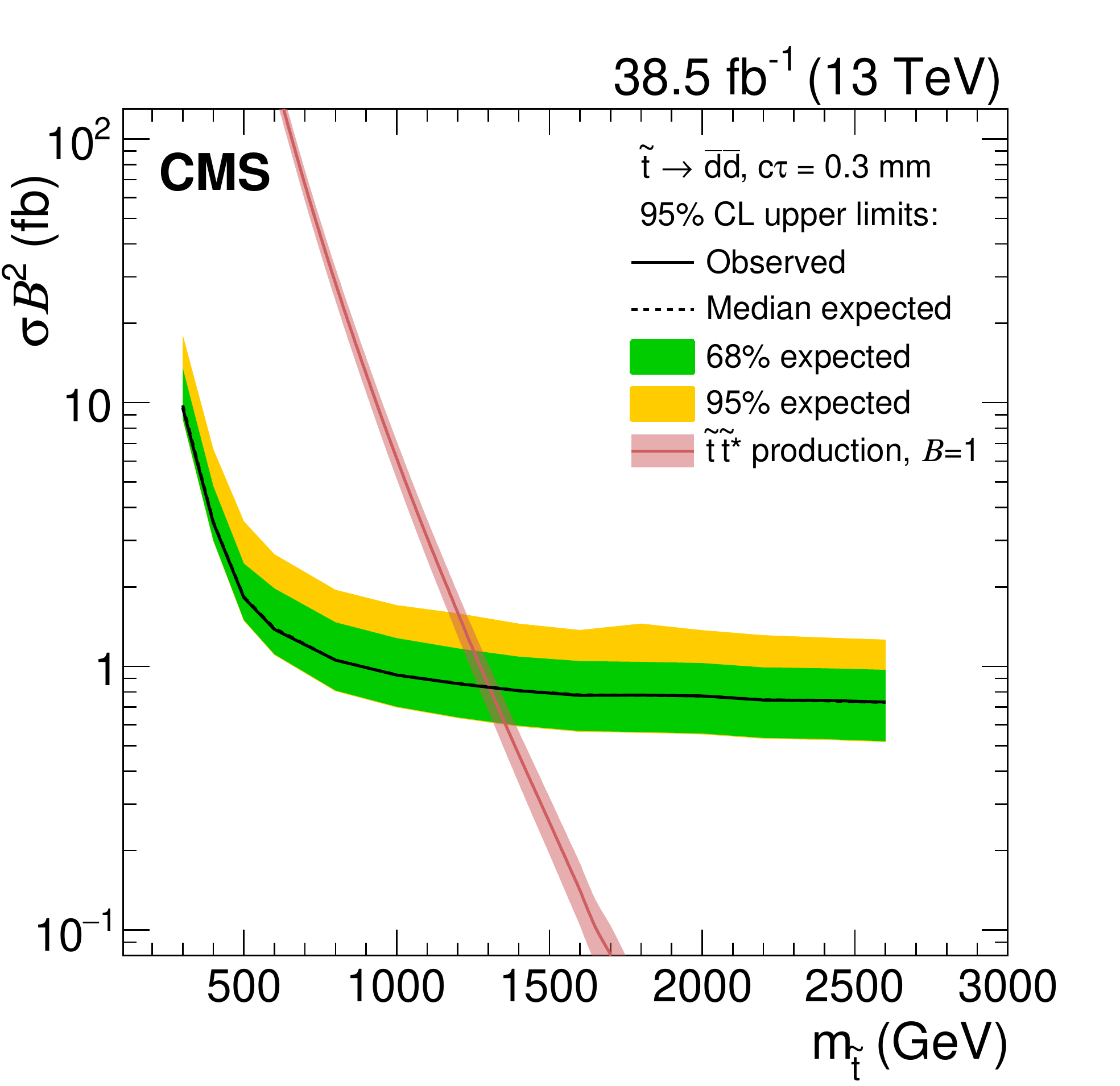}
\includegraphics[width=0.45\textwidth]{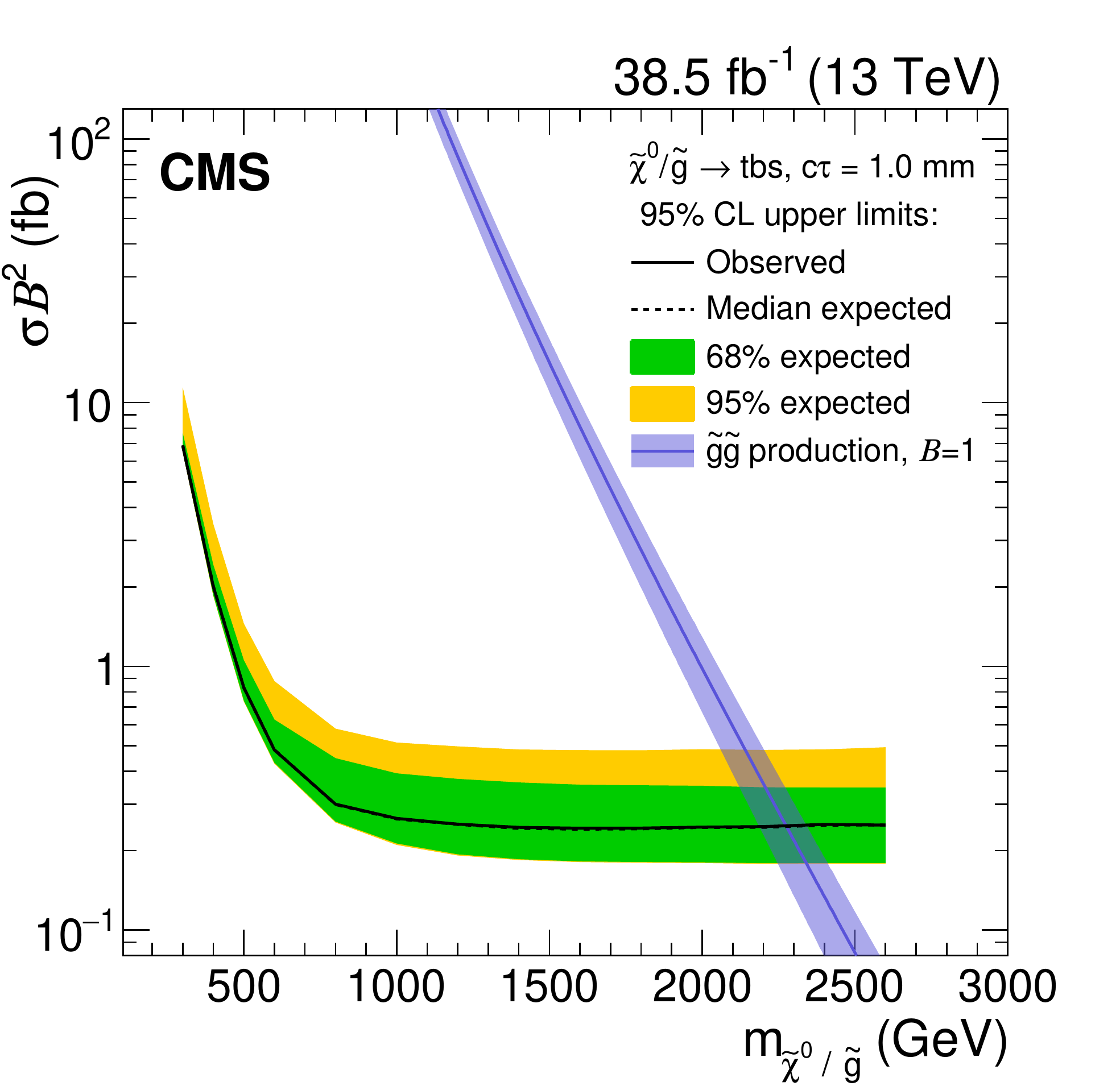}
\includegraphics[width=0.45\textwidth]{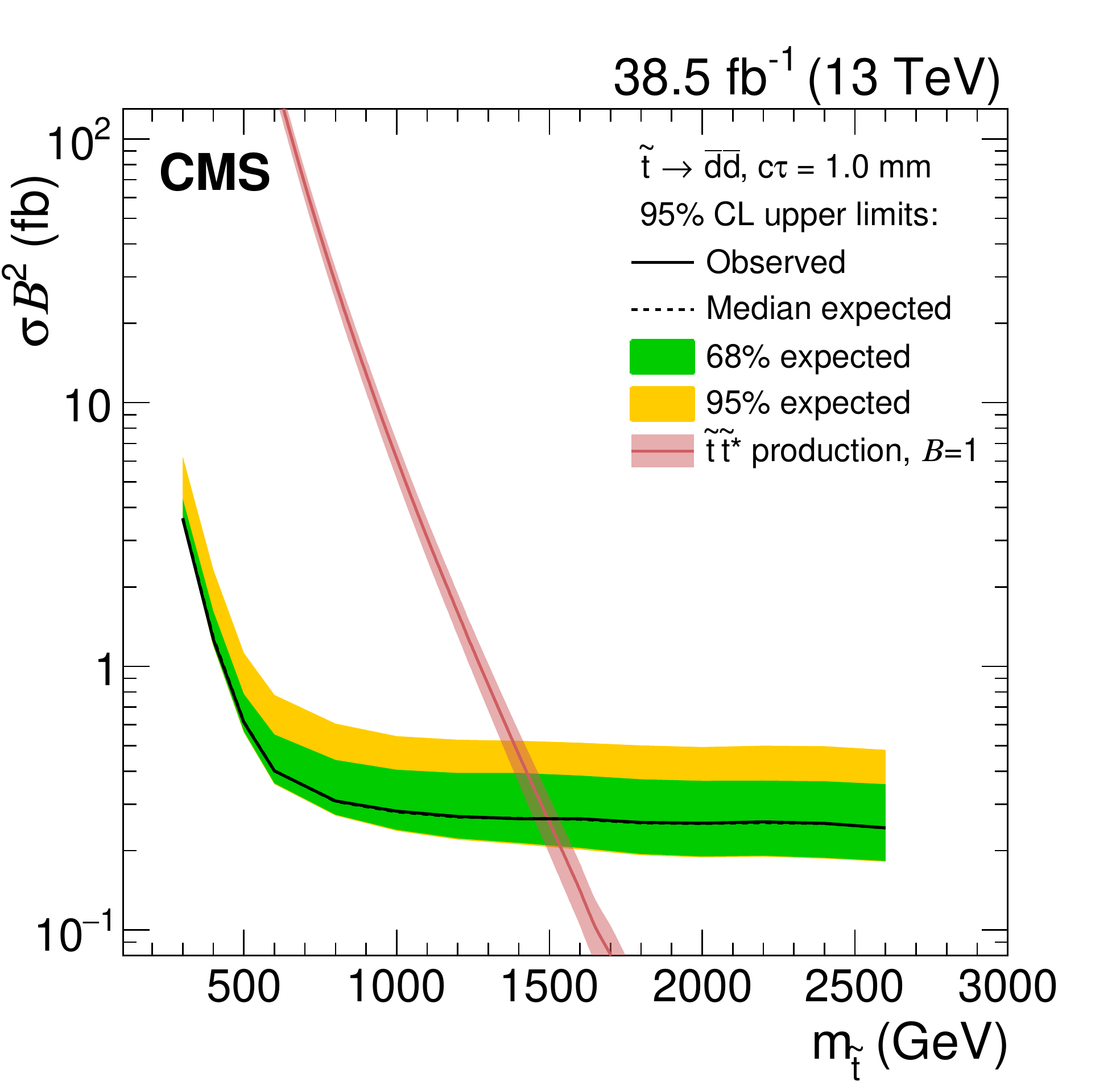}
\includegraphics[width=0.45\textwidth]{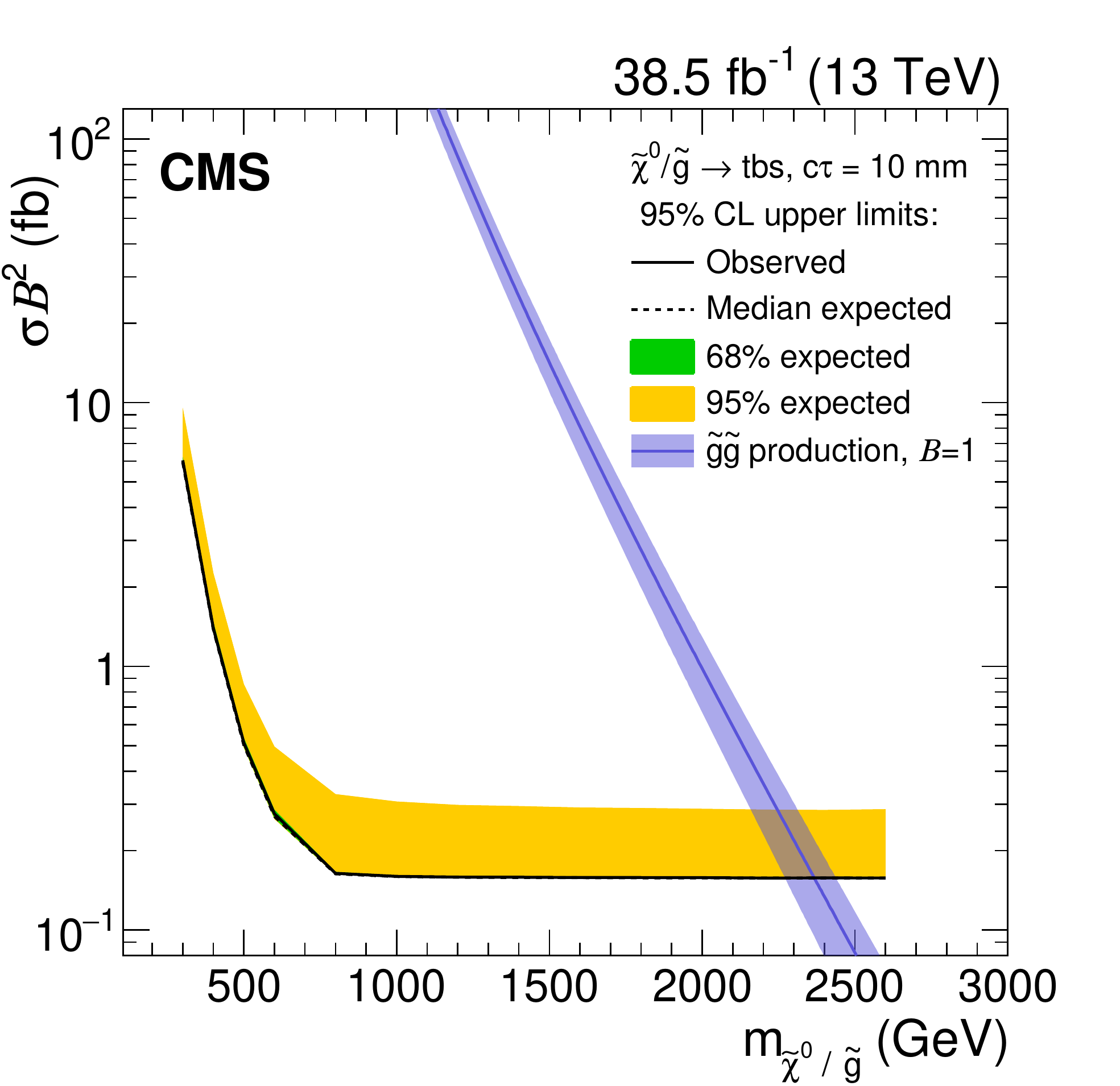}
\includegraphics[width=0.45\textwidth]{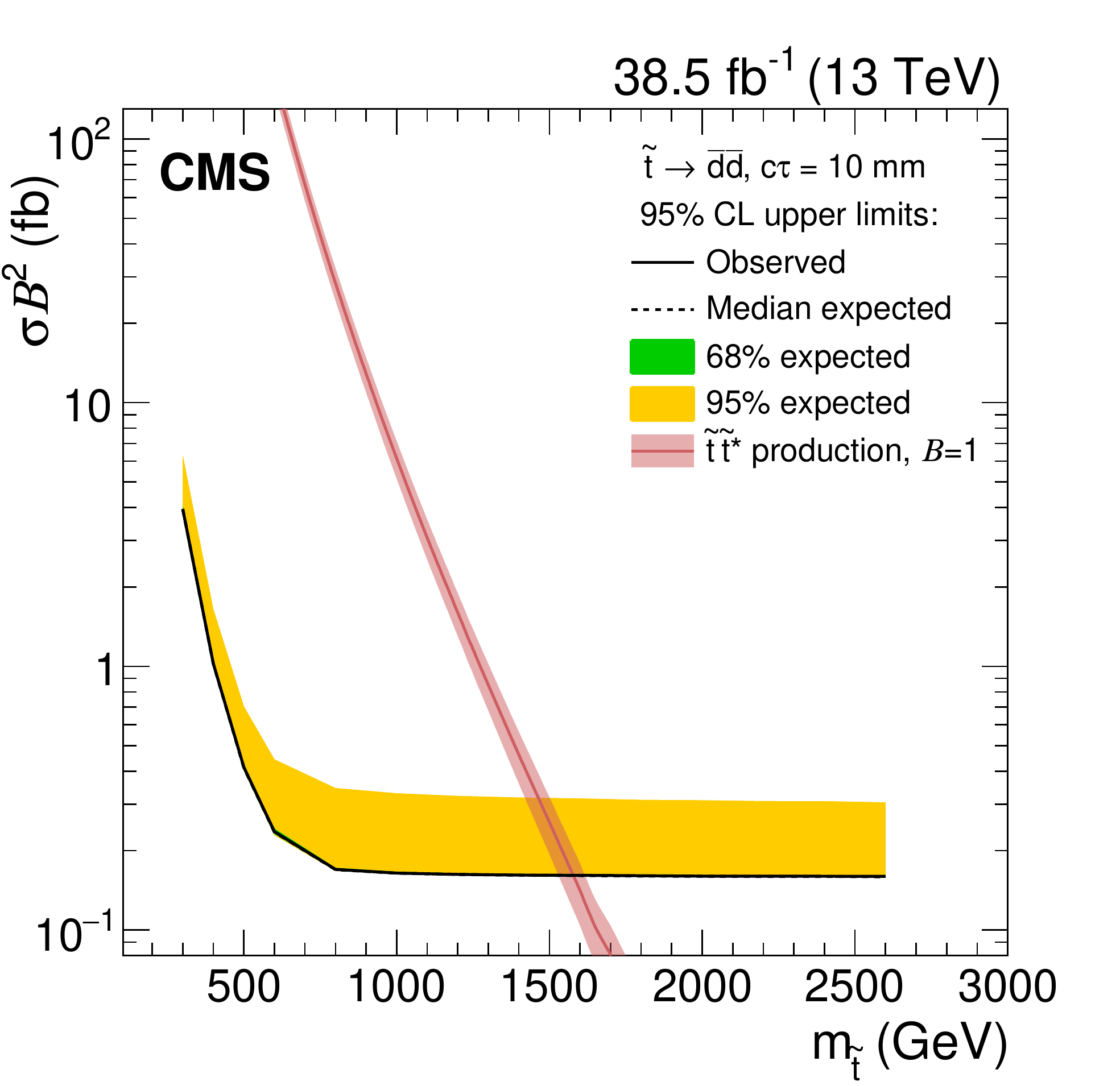}
\caption{Observed and expected 95\% \CL upper limits on
$\sigma\mathcal{B}^2$ for the multijet (left) and dijet (right)
signals, as a function of mass for a fixed $c\tau$ of 0.3\mm
(upper), 1.0\mm (middle), and 10\mm (lower).  The gluino pair
production cross section is overlaid for the multijet signals, and
the top squark pair production cross section is overlaid for the
dijet signals.  The uncertainties in the theoretical cross sections
include those due to the renormalization and factorization scales,
and the parton distribution functions.}
\label{fig:limits1d_vs_mass}
\end{figure*}

\begin{figure*}[hbtp]
\centering
\includegraphics[width=0.45\textwidth]{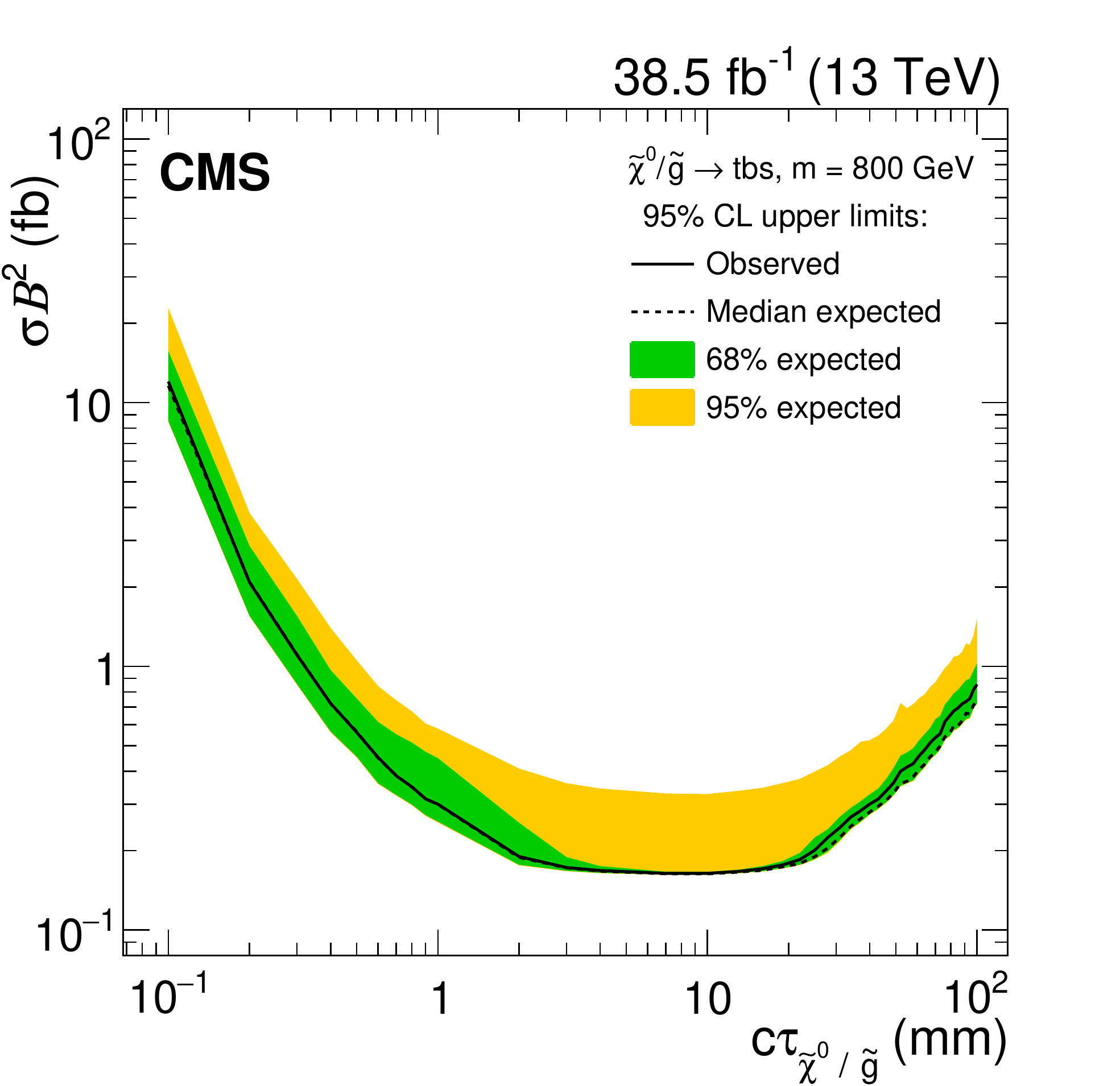}
\includegraphics[width=0.45\textwidth]{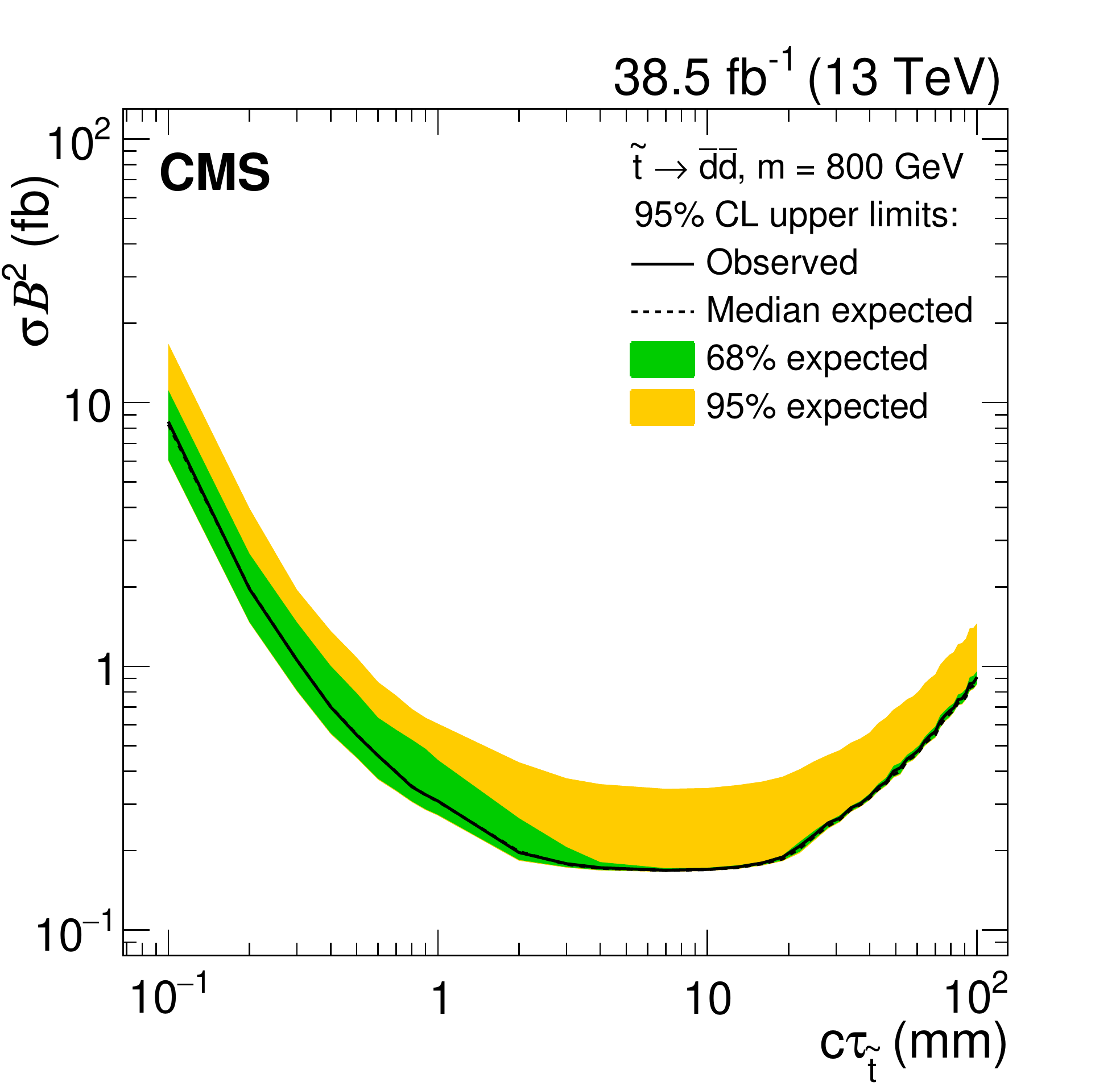}
\includegraphics[width=0.45\textwidth]{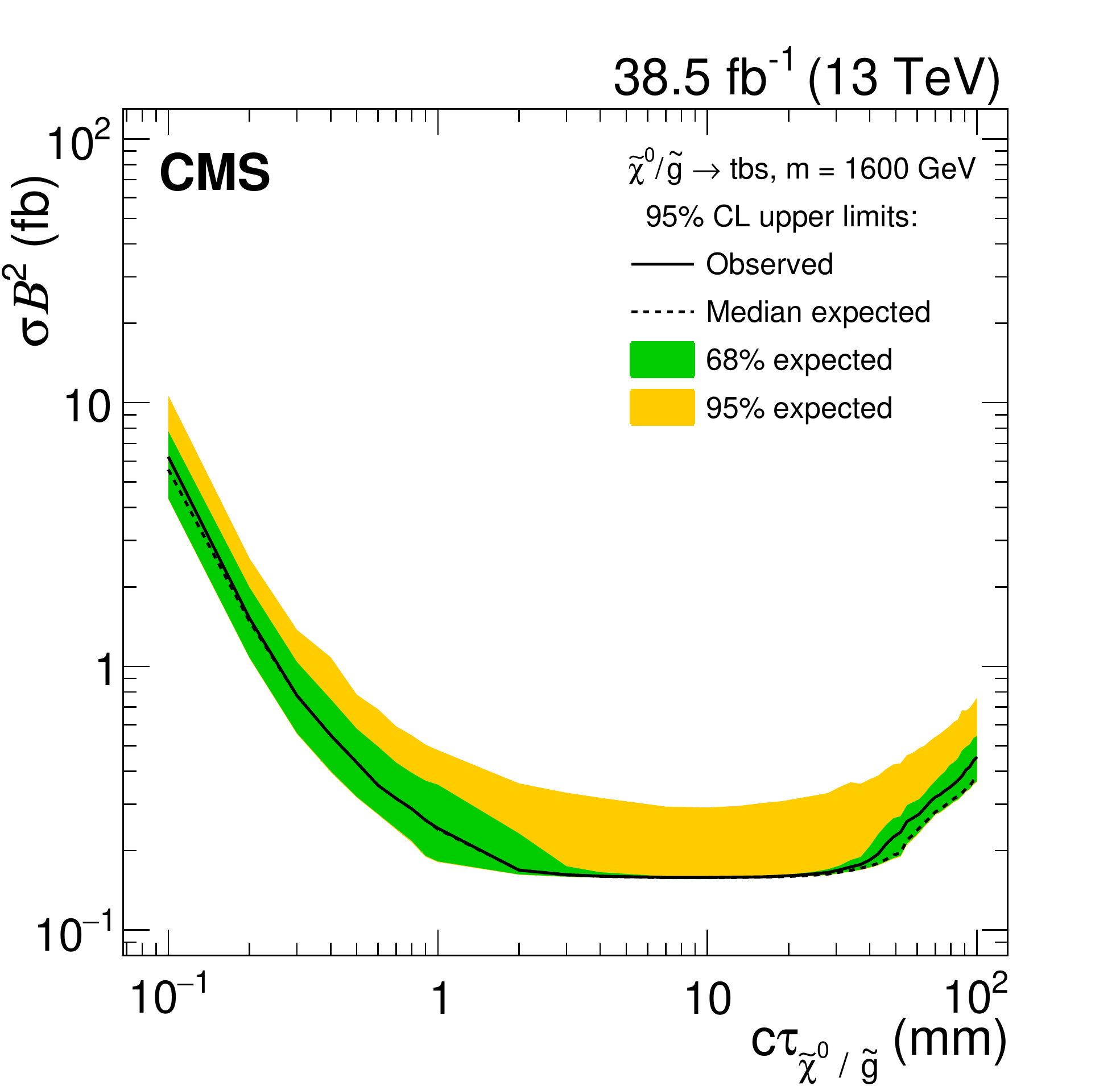}
\includegraphics[width=0.45\textwidth]{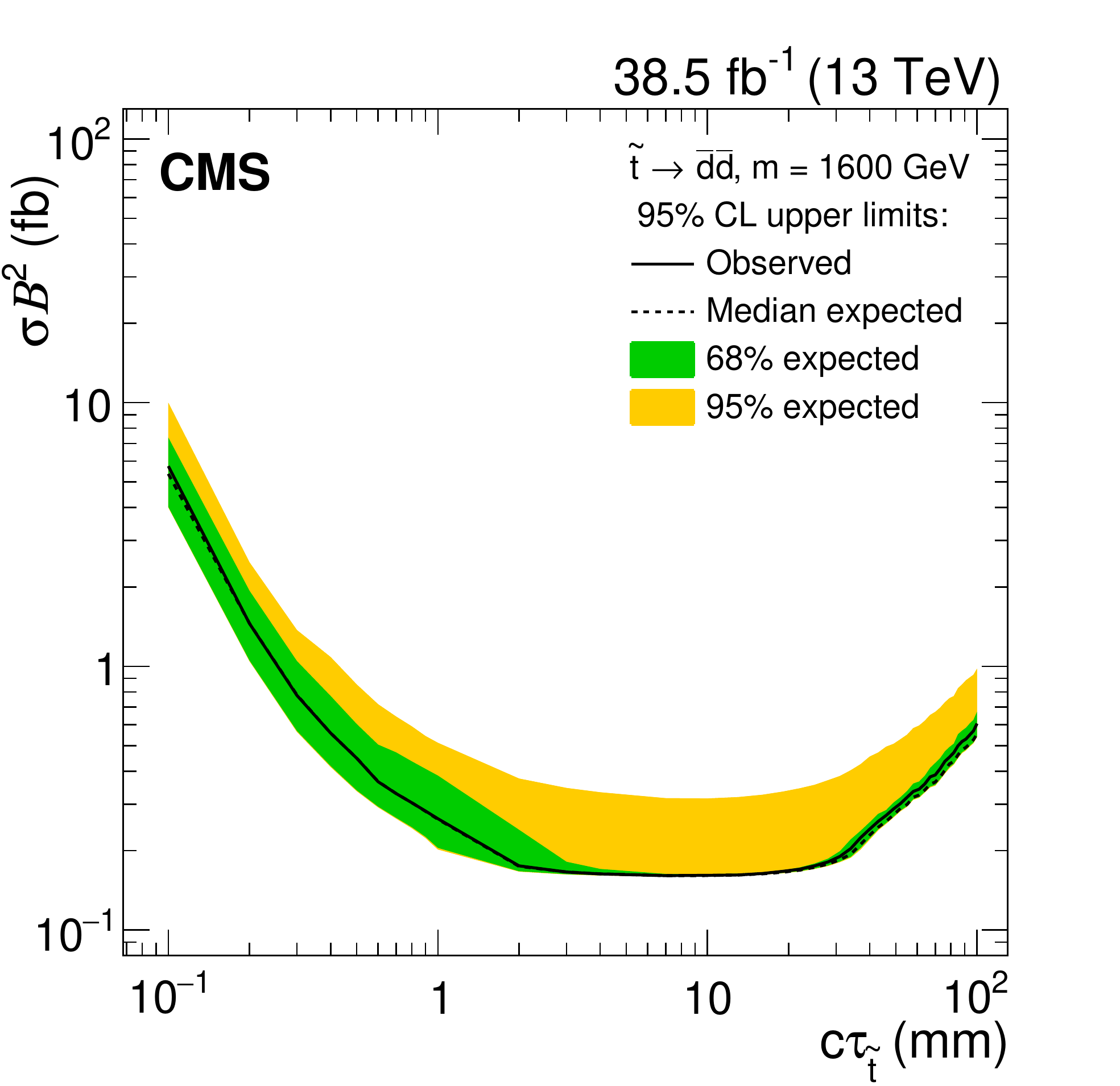}
\includegraphics[width=0.45\textwidth]{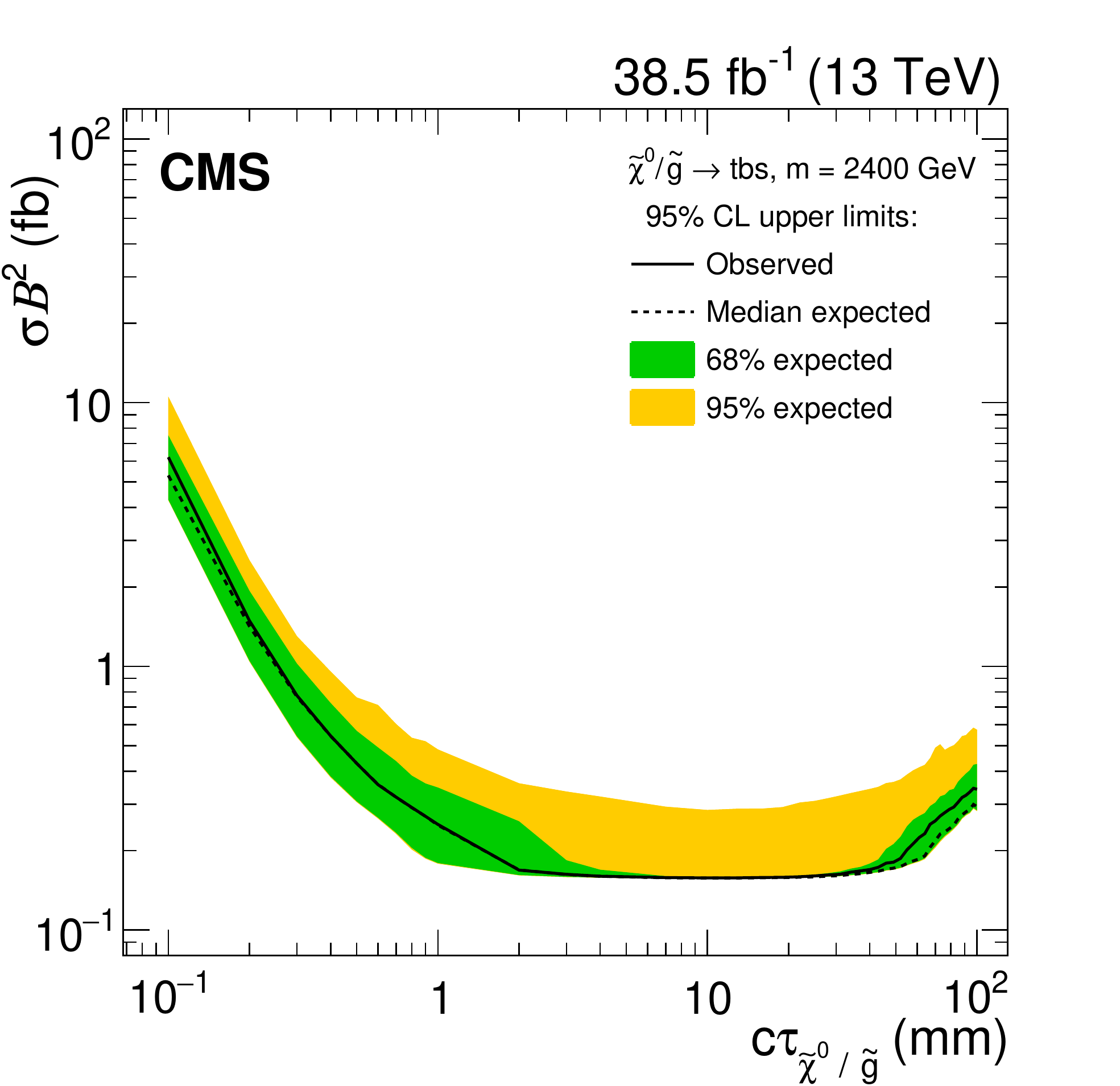}
\includegraphics[width=0.45\textwidth]{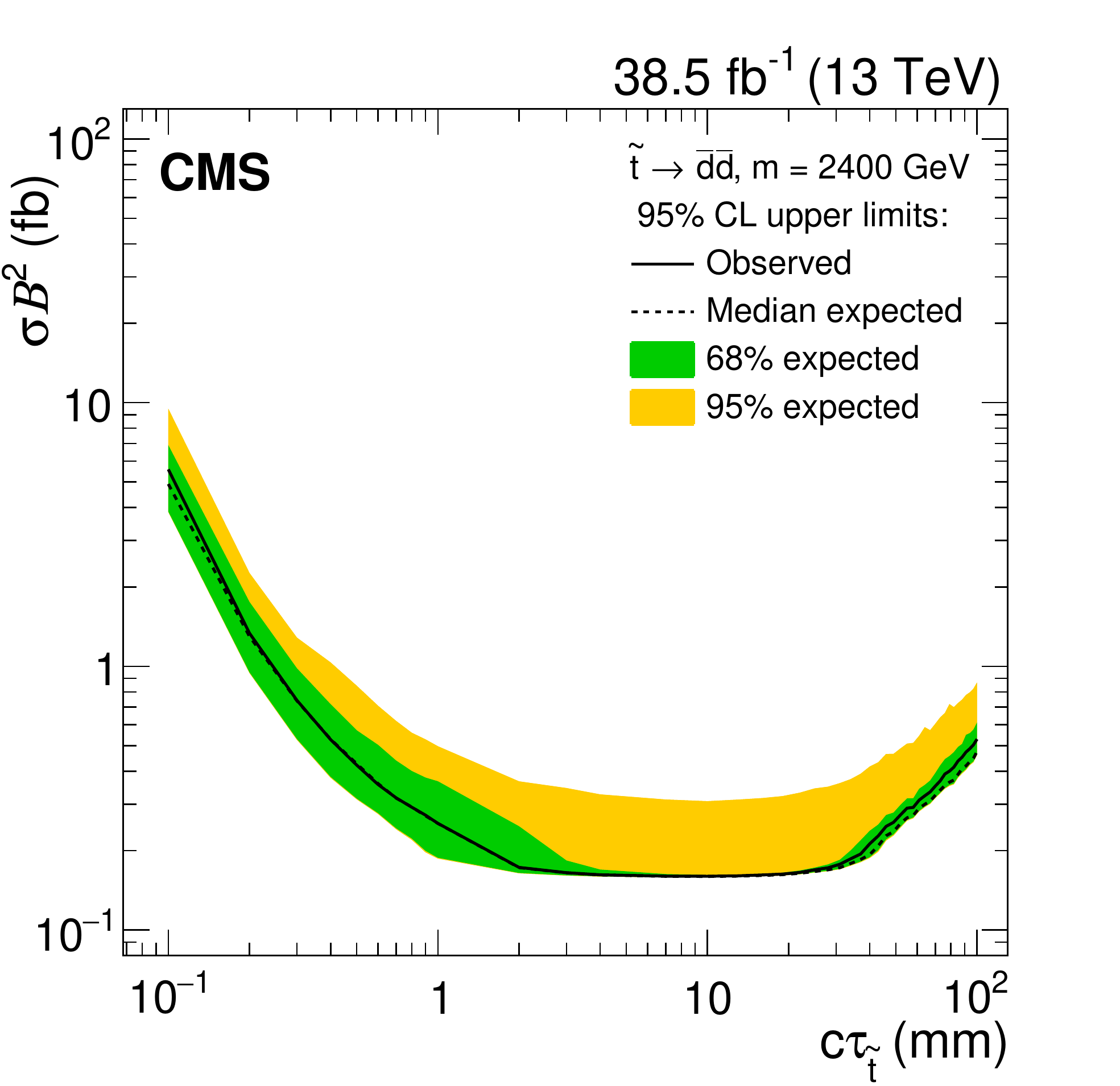}
\caption{Observed and expected 95\% \CL upper limits on
$\sigma\mathcal{B}^2$ for the multijet (left) and dijet (right)
signals, as a function of $c\tau$ for a fixed mass of 800\GeV
(upper), 1600\GeV (middle), and 2400\GeV (lower).}
\label{fig:limits1d_vs_ctau}
\end{figure*}

Figure~\ref{fig:limits} shows, as a function of lifetime and mass,
the observed 95\% confidence level (\CL) upper limits on the product
of the signal pair production cross section and the square of the
branching fraction for its decay ($\sigma\mathcal{B}^2$) for both
the multijet and dijet signals.  The expected limits are similar.
Exclusion curves are overlaid, assuming gluino and top squark pair
production cross sections~\cite{gluinoxsec} and 100\% branching
fraction, for both the observed and expected 95\% \CL upper limits.
The upper limits reflect the signal efficiencies shown in
Fig.~\ref{fig:sigeff}, initially improving as lifetime increases,
but worsening at approximately 40\mm due to the fiducial limit at
the beam pipe.  As an example, for a neutralino with mass of 800\GeV
and $c\tau = 1\mm$, the observed 95\% \CL upper limit on
$\sigma\mathcal{B}^2$ is 0.3\unit{fb}.  For mean proper decay
lengths between 0.6 and 80\mm, gluino masses are excluded below
2200\GeV, and top squark masses are excluded below 1400\GeV.
Figure~\ref{fig:limits1d_vs_mass} shows the upper limits as a
function of mass for several values of $c\tau$, and
Fig.~\ref{fig:limits1d_vs_ctau} shows the upper limits as a function
of $c\tau$ for several values of the mass.

In Fig.~\ref{fig:limits1d_vs_ctau}, the narrowing of the expected
limit bands above $c\tau = 2\mm$ is due to the correlation between
the signal lifetime and the relative signal yields in the three \dvv
bins.  The low background yield causes the discrete nature of the
Poisson distribution to have an effect: the pseudodata sets used to
calculate the distribution of expected limits have a limited number
of combinations of yields in each bin.  For example, for a simulated
multijet signal with $m = 1600\GeV$ and $c\tau = 4\mm$, the signal
is concentrated almost entirely ($>$90\%) in the last bin.  The
majority of pseudodata sets that are different in only the first
two bins then have nearly the same expected limit value.  The bands
widen above $c\tau = 20\mm$ with the reappearance of signal in the
first bin due to the effect described in
Section~\ref{sec:searchstrategy} in which two vertices are
reconstructed from the same long-lived particle, an effect that is
larger for the multijet signals.

\section{Extending the search to other signal models}

This search for displaced vertices applies to other types of
long-lived particles decaying to multiple jets.  Here we present a
generator-level selection that can be used to reinterpret the
results of our analysis.  For signal models in which there are two
long-lived particles, this generator-level selection approximately
replicates the reconstruction-level efficiency.  The selection is
based on the number and momenta of generated jets in the event, the
displacements of the long-lived particles, and the momenta of their
daughter particles.  The generated jets are those clustered from all
final-state particles except neutrinos, using the anti-\kt algorithm
with a distance parameter of 0.4, but are rejected if the fraction
of energy from electrons is greater than 0.9 or if the fraction of
energy from muons is greater than 0.8.  The daughter particles are
the \PQu, \PQd, \PQs, \PQc, and \PQb quarks, electrons, muons, and
tau leptons from the decay of the long-lived particle, and we
consider those with an impact parameter with respect to the origin
measured in the $x$-$y$ plane of at least 0.1\mm.  The generated
jets and daughter particles are required to satisfy $\pt > 20\GeV$
and $\abs{\eta} < 2.5$.

The criteria of the generator-level selection are as follows: at
least four generated jets; $\HT > 1000\GeV$, where \HT is the scalar
sum of the \pt of generated jets with $\pt > 40\GeV$; for each
long-lived particle, a distance of the decay point from the origin
measured in the $x$-$y$ plane of between 0.1 and 20\mm, and a value
of $\Sigma\pt$ of the daughter particles of at least 350\GeV; and a
distance between the decay points of the long-lived particles
measured in the $x$-$y$ plane of at least 0.4\mm.  In calculating
the $\Sigma\pt$ of the daughter particles, we multiply the \pt of
\PQb quark daughter particles by a factor of 0.65.  This accounts
for the lower reconstruction-level efficiency due to the lifetime of
heavy flavor particles, which can impede the association of their
decay products with the reconstructed vertices.

This generator-level selection replicates the reconstruction-level
efficiency with a typical accuracy of 20\% for a variety of models
for which the signal efficiency is high ($>$10\%).  In the region
with $\dvv > 0.4\mm$, there are no observed events.

\section{Summary}

A search for long-lived particles decaying into multijet final
states has been performed using proton-proton collision events
collected with the CMS detector at a center-of-mass energy of 13\TeV
in 2015 and 2016.  The data sample corresponds to an integrated
luminosity of \intlumiTotal.  No excess yield above the prediction
from standard model processes is observed.  At 95\% confidence
level, upper limits are placed for models of $R$-parity violating
supersymmetry in which the long-lived particles are neutralinos or
gluinos decaying solely into multijet final states or top squarks
decaying solely into dijet final states.  The data exclude cross
sections above approximately 0.3\unit{fb} for particles with masses
between 800 and 2600\GeV and mean proper decay lengths between 1 and
40\mm.  For mean proper decay lengths between 0.6 and 80\mm, gluino
masses below 2200\GeV and top squark masses below 1400\GeV are
excluded.  While the search specifically addresses two models of
$R$-parity violating supersymmetry, the results are relevant to
other models in which long-lived particles decay to final states
with multiple tracks, and a method to extend the search to other
signal models is provided.  For the models considered, the results
provide the most restrictive bounds to date on the production and
decay of pairs of long-lived particles with mean proper decay
lengths between 0.1 and 100\mm.

\begin{acknowledgments}
We congratulate our colleagues in the CERN accelerator departments for the excellent performance of the LHC and thank the technical and administrative staffs at CERN and at other CMS institutes for their contributions to the success of the CMS effort. In addition, we gratefully acknowledge the computing centers and personnel of the Worldwide LHC Computing Grid for delivering so effectively the computing infrastructure essential to our analyses. Finally, we acknowledge the enduring support for the construction and operation of the LHC and the CMS detector provided by the following funding agencies: BMWFW and FWF (Austria); FNRS and FWO (Belgium); CNPq, CAPES, FAPERJ, and FAPESP (Brazil); MES (Bulgaria); CERN; CAS, MoST, and NSFC (China); COLCIENCIAS (Colombia); MSES and CSF (Croatia); RPF (Cyprus); SENESCYT (Ecuador); MoER, ERC IUT, and ERDF (Estonia); Academy of Finland, MEC, and HIP (Finland); CEA and CNRS/IN2P3 (France); BMBF, DFG, and HGF (Germany); GSRT (Greece); NKFIA (Hungary); DAE and DST (India); IPM (Iran); SFI (Ireland); INFN (Italy); MSIP and NRF (Republic of Korea); LAS (Lithuania); MOE and UM (Malaysia); BUAP, CINVESTAV, CONACYT, LNS, SEP, and UASLP-FAI (Mexico); MBIE (New Zealand); PAEC (Pakistan); MSHE and NSC (Poland); FCT (Portugal); JINR (Dubna); MON, RosAtom, RAS and RFBR (Russia); MESTD (Serbia); SEIDI, CPAN, PCTI and FEDER (Spain); Swiss Funding Agencies (Switzerland); MST (Taipei); ThEPCenter, IPST, STAR, and NSTDA (Thailand); TUBITAK and TAEK (Turkey); NASU and SFFR (Ukraine); STFC (United Kingdom); DOE and NSF (USA).

\hyphenation{Rachada-pisek} Individuals have received support from the Marie-Curie program and the European Research Council and Horizon 2020 Grant, contract No. 675440 (European Union); the Leventis Foundation; the A. P. Sloan Foundation; the Alexander von Humboldt Foundation; the Belgian Federal Science Policy Office; the Fonds pour la Formation \`a la Recherche dans l'Industrie et dans l'Agriculture (FRIA-Belgium); the Agentschap voor Innovatie door Wetenschap en Technologie (IWT-Belgium); the F.R.S.-FNRS and FWO (Belgium) under the ``Excellence of Science - EOS" - be.h project n. 30820817; the Ministry of Education, Youth and Sports (MEYS) of the Czech Republic; the Lend\"ulet (``Momentum") Programme and the J\'anos Bolyai Research Scholarship of the Hungarian Academy of Sciences, the New National Excellence Program \'UNKP, the NKFIA research grants 123842, 123959, 124845, 124850 and 125105 (Hungary); the Council of Science and Industrial Research, India; the HOMING PLUS program of the Foundation for Polish Science, cofinanced from European Union, Regional Development Fund, the Mobility Plus program of the Ministry of Science and Higher Education, the National Science Center (Poland), contracts Harmonia 2014/14/M/ST2/00428, Opus 2014/13/B/ST2/02543, 2014/15/B/ST2/03998, and 2015/19/B/ST2/02861, Sonata-bis 2012/07/E/ST2/01406; the National Priorities Research Program by Qatar National Research Fund; the Programa Estatal de Fomento de la Investigaci{\'o}n Cient{\'i}fica y T{\'e}cnica de Excelencia Mar\'{\i}a de Maeztu, grant MDM-2015-0509 and the Programa Severo Ochoa del Principado de Asturias; the Thalis and Aristeia programs cofinanced by EU-ESF and the Greek NSRF; the Rachadapisek Sompot Fund for Postdoctoral Fellowship, Chulalongkorn University and the Chulalongkorn Academic into Its 2nd Century Project Advancement Project (Thailand); the Welch Foundation, contract C-1845; and the Weston Havens Foundation (USA).
\end{acknowledgments}

\bibliography{auto_generated}
\cleardoublepage \appendix\section{The CMS Collaboration \label{app:collab}}\begin{sloppypar}\hyphenpenalty=5000\widowpenalty=500\clubpenalty=5000\vskip\cmsinstskip
\textbf{Yerevan Physics Institute, Yerevan, Armenia}\\*[0pt]
A.M.~Sirunyan, A.~Tumasyan
\vskip\cmsinstskip
\textbf{Institut f\"{u}r Hochenergiephysik, Wien, Austria}\\*[0pt]
W.~Adam, F.~Ambrogi, E.~Asilar, T.~Bergauer, J.~Brandstetter, M.~Dragicevic, J.~Er\"{o}, A.~Escalante~Del~Valle, M.~Flechl, R.~Fr\"{u}hwirth\cmsAuthorMark{1}, V.M.~Ghete, J.~Hrubec, M.~Jeitler\cmsAuthorMark{1}, N.~Krammer, I.~Kr\"{a}tschmer, D.~Liko, T.~Madlener, I.~Mikulec, N.~Rad, H.~Rohringer, J.~Schieck\cmsAuthorMark{1}, R.~Sch\"{o}fbeck, M.~Spanring, D.~Spitzbart, A.~Taurok, W.~Waltenberger, J.~Wittmann, C.-E.~Wulz\cmsAuthorMark{1}, M.~Zarucki
% ...
\vskip\cmsinstskip
% ...
1:  Also at Vienna University of Technology, Vienna, Austria\\
% ...
\end{sloppypar}
\end{document}